 \documentclass[10pt,preprint]{aastex}  %compact preprint style (MJA)
 \usepackage{natbib}
\def\ang{\AA}
\def\arcsec{\hbox{$^{\prime\prime}$}}

\def\gapprox{\lower.4ex\hbox{$\;\buildrel >\over{\scriptstyle\sim}\;$}}
\def\lapprox{\lower.4ex\hbox{$\;\buildrel <\over{\scriptstyle\sim}\;$}}

\shortauthors{ASCHWANDEN}
\shorttitle{Torsional Oscillations in Solar Flares}

\begin{document}

\title{ Torsional Alfv\'enic Oscillations Discovered in the Magnetic 
	Free Energy During Solar Flares}

\author{Markus J. Aschwanden}

\affil{	Solar and Astrophysics Laboratory,
	Lockheed Martin Advanced Technology Center, 
        Dept. ADBS, Bldg.252, 3251 Hanover St., Palo Alto, CA 94304, USA; 
        (e-mail: \url{aschwanden@lmsal.com})}

\and

\author{Tongjiang Wang }

\affil{ The Catholic University of America and 
	NASA Goddard Space Flight Center,
	Code 671, Greenbelt, MD 20770, USA
	(e-mail: \url{tongjiang.wang@nasa.gov})} 

\begin{abstract}
We report the discovery of torsional Alfv\'enic oscillations in 
solar flares, which modulate the time evolution of the magnetic free 
energy $E_f(t)$, while the magnetic potential energy $E_p(t)$ is 
uncorrelated, and the nonpotential energy varies as 
$E_{np}(t) = E_p + E_f(t)$. The mean observed time period of the 
torsional oscillations is $P_{obs}=15.1 \pm 3.9$ min, the mean field
line length is $L=135\pm35$ Mm, and the mean phase speed is $v_{phase}
=315 \pm 120$ km s$^{-1}$, which we interpret as torsional
Alfv\'enic waves in flare loops with enhanced electron densities.
Most of the torsional oscillations are found to be decay-less,
but exhibit a positive or negative trend in the evolution of the
free energy, indicating new emerging flux (if positive),
magnetic cancellation, or flare energy dissipation (if negative).
The time evolution of the free energy has been calculated
in this study with the {\sl Vertical-Current Approximation 
(Version 4) Nonlinear Force-Free Field (VCA4-NLFFF)} code,
which incorporates automatically detected coronal loops in
the solution and bypasses the non-forcefreeness of the 
photospheric boundary condition, in contrast to traditional
NLFFF codes. 
\end{abstract}


\section{	INTRODUCTION 				}

Torsional Oscillations have been explored first in classical 
physics with the so-called {\sl ``torsional pendulum''}, where a 
mass is suspended with a wire, which rotates first in 
one circular direction, and then in the reverse direction,
while the torque in the wire represents the restoring force. 

Today, various types of torsional oscillations have been studied 
in solar physics, in the context of: 
(i) the solar dynamo that oscillates between the global poloidal and
toroidal magnetic field in an 11-year solar cycle (Howard and LaBonte 1980;
Kitchatinov et al.~1999; Durney 2000; Chakraborty et al.~2009;
Gurerrero et al.~2016; Lekshmi et al.~2018; Kosovichev and Pipin 2019) 
or stellar cycle (Lanza 2007);
(ii) the differential solar rotation, where zonal latitude bands on 
the solar surface exhibit slower- and faster-than average rotation speeds
(Spruit 2003; Rempel 2006, 2007, 2012; Howe et al.~2006, 2009, 2013); 
(iii) meridional circulation (Gonzalez-Hernandez et al.~2010);
(iv) torsional oscillations in the solar (or stellar) convection zone
(Covas et al.~2000, 2004; Noble et al.~2003; Zhao and Kosovichev 2004); 
(v) helioseismic measurements of torsional oscillations inside
the Sun, using time-distance helioseimology measurements and inversions 
of {\sl Michelson Doppler Imager (MDI)} data onboard the
{\sl Solar and Heliospheric Observatory (SoHO)} spacecraft,
MDI/SOHO data (Zhao and Kosovichev 2004; Vorontsov et al.~2002), 
or ring-diagram analysis of {\sl Global Oscillation Network Group
(GONG)} data (Howe et al.~2006; Lekshmi et al.~2018); 
(vi) torsional and compressional waves in photospheric flux tubes
(Sakai et al.~2001; Luo et al.~2002; Routh et al.~2007, 2010)
and intergranular magnetic flux concentrations (Shelyag 2015); 
(vii) torsional oscillations of sunspots (Gopasyuk and Kosovichev 2011;
Grinon-Marin et al.~2017), 
(viii) torsional oscillations in the generation of active regions
(Petrovay and Forgacs-Dajka 2002);
(viii) torsional oscillations of coronal loops or threads
(Zaqarashvili et al.~2003, 2013; Zaqarashvili and Murawski 2007;
Zaqarashvili and Belvedere 2007; Copil et al.~2008, 2010;
Vasheghani-Farahani et al. 2010, 2011; 
Mozafari Ghoraba and Vasheghani-Farahani 2018); 
(ix) ubiquitous torsional motions in type II spicules 
(De Pontieu et al.~2012; Sekse et al.~2013);
(x) Torsional wave propagation in solar tornados 
(Vasheghani-Farahani et al.~2017); 
(xi) a torsional wave model for solar radio pulsations
(Tapping 1983); 
(xii) torsional Alfv\'en waves embedded in small magnetic
flux ropes in the solar wind (Gosling et al.~2010;
Higginson and Lynch 2018) and
in interplanetary magnetic clouds (Raghav et al.~2018;
Raghav and Kule 2018).
Torsional oscillations have also been invoked in other fields 
of astrophysics, such as in a hyperflare from a soft gamma-ray
repeater (Strohmayer and Watts 2006), or torsional oscillations in
a magnetar (Link and van Eysden 2016).  

Theoretical studies on torsional oscillations in the solar plasma
include: 
modeling the Lorentz force and angular
momentum associated with the mean field dynamo model
(Durney 2000; Covas et al.~2000; Lanza 2007); 
torsional Alfv\'en waves and their role in coronal heating 
(Narain et al.~2001; Antolin and Shibata 2010); 
MHD simulations of torsional and compressional waves in 
photospheric flux tubes
(Sakai et al.~2001; Luo et al.~2002; Routh et al.~2007, 2010);
analytical models of geostrophic flows (Spruit 2003)
and turbulent flows in the convection zone (Noble et al.~2003); 
flux-transport dynamo models with Lorentz force feedback
and quenching of meridional flows by turbulent viscosity
and heat conductivity (Rempel 2006); 
MHD simulations that reproduce the latitudinal rotation rates 
(Guerrero et al.~2016); 
eigenmodes of torsional Alfv\'en waves in stratified solar
waveguides (Verth et al.~2010; Karami and Bahari 2011); 
frequency filtering of torsional Alfv\'en waves (Fedun et al.~2011); 
magneto-seismology in static and dynamic coronal plasmas
(Morton et al.~2011);
partial ionization, neutral helium, and stratification effects
(Zaqarashvili et al.~2013); 
MHD simulations of torsional Alfv\'en
waves in flux tubes with axial symmetry (Muraski et al.~2015;
Wojcik et al.~2017); 
propagation of torsional Alfv\'en waves
in expanding flux tubes (Soler et al.~2017; 2019); 
or magnetic shocks excited by torsional Alfv\'en wave interactions 
(Snow et al.~2018).

The particular technique used in this paper to detect 
and measure torsional oscillations in the solar corona is a new 
approach, based on automated tracing of coronal loop structures 
and on forward-fitting of analytical solutions of helically
twisted loops in a non-linear force-free field, using
line-of-sight magnetograms (from the {\sl Helioseismic and 
Magnetic Imager (HMI)} onboard the {\sl Solar Dynamic Observatory
(SDO)} (Scherrer et al.~2012) and {\sl extreme ultra-violet (EUV)}
images from the {\sl Atmospheric Imager Assembly (AIA)} 
(Lemen et al.~2012)
onboard SDO. The parameterization of the analytical
magnetic field model yields directly the time evolution of 
helically twisted (azimuthal) magnetic field components 
$B_{\varphi}({\bf x},t)$ at location ${\bf x}$ and time $t$,
which is an ideal parameter to track torsional oscillations.
Once we improved the accuracy of the {\sl Vertical-Current
Approximation Non-Linear Force-Free Field (VCA4-NLFFF)} 
solutions with the latest version 4 of this code, we 
discovered torsional oscillations in all analyzed (GOES X-class) 
solar flares.

The contents of this paper include the theoretical concept
(Section 2), observations and analysis of AIA and
HMI data (Section 3), results (Section 4), discussion and 
interpretation (Section 5), conclusions (Section 6), and a 
brief description of the updated VCA4-NLFFF code 
(Appendix A).

\section{	THEORETICAL CONCEPT 			}

\subsection{	The Free Energy				}

This study is devoted to understand a new type of torsional
oscillations that has been discovered during solar 
flares and is described and modeled here for the first time. 
This discovery has not been predicted for solar flares and
emerged as a spin-off from earlier studies on the global 
energetics of solar flares (Aschwanden et al.~2014a;
Aschwanden 2019c), where we attempted to measure the  
time evolution of the (magnetic) free energy 
$E_{f}(t)$, using a {\sl non-linear force-free field 
(NLFFF)} code. The free energy $E_{f}$ 
can be derived from the potential field ${\bf B}_{p}$ and
the nonpotential field solution ${\bf B}_{np}$, being
the difference between these two {\bf volume-integrated} quantities,
\begin{equation}
	E_{f} = E_{np} - E_p =
	\int {{\bf B}_{np}^2({\bf r})\over 8 \pi}\ dV -  
	\int {{\bf B}_{p}^2({\bf r})\over 8 \pi}\ dV =
	\int {{\bf B}_{f}^2({\bf r})\over 8 \pi}\ dV \ ,
\end{equation}
where the free energy $B_{f}({\bf r})$ is given here by the azimuthal 
magnetic field component, i.e., $B_f({\bf r})= B_\varphi({\bf r})$, 
in the framework of spherical coordinates 
${\bf B}=(B_r, B_{\varphi}, B_{\theta})$. 
The integration over
the 3-D volume renders the total magnetic energy in a flare or an 
active region. The geometric concept of a torsionally twisted flux
tube is depicted in Fig.~1. Eq.~(1) reminds us of the Pythagorean 
geometric relationship in an orthogonal 
triangle with sides $B_{np}$, $B_p$, and $B_{\varphi}$.
The radial (potential) field direction $B_p$ is orthogonal
to the azimuthal (twisting) field component $B_{\varphi}$, 
as visualized in Fig.~1, so that the sum of the squared 
sides yield,
\begin{equation}
	B_{np}^2({\bf r)} = B_p^2({\bf r}) + B_\varphi^2({\bf r}) \ .
\end{equation}
consistent with the sum of energies in Eq.~(1). 

We can now parameterize torsional oscillations in a
helically twisted (straight) flux tube. 
In this geometric concept, the potential field and the
azimuthal component can always be decoupled, and thus
the potential field can be kept stationary, while the
azimuthal component can oscillate independently, so that
the vector Eq.~(2) reads as,
\begin{equation}
	E_{np}(t) = E_p + E_\varphi(t) \ .
\end{equation}
We can model the time variation of the free energy 
of a torsionally twisted flux tube with a sine function,
\begin{equation}
	E_{f}(t) = E_{\varphi}(t) = 
	E_{f0} \  \sin{ \left( {2 \pi (t-t_0) \over P}
	\right) } \ ,
\end{equation}
where the time evolution of the free energy $E_{f}(t)$ 
or $E_\varphi(t)$ is a direct observable in our data analysis.

\subsection{	The Magnetic Field 			}

We use a {\sl nonlinear force-free field (NLFFF)} code
that satisfies the divergence-free condition
\begin{equation}
        \nabla \cdot {\bf B} = 0 \ ,
\end{equation}
and the force-freeness condition,
\begin{equation}
        \nabla \times {\bf B} = \alpha({\bf r}) {\bf B} 
 	= 4 \pi {\bf j} \ ,
\end{equation}
where $\alpha({\bf r})$ represents a scalar function that depends
on the position ${\bf r}$, but is constant along a magnetic field
line, and ${\bf j}$ represents the current density.
Three different types of magnetic fields are generally
considered in applications to the solar corona: (i) a {\sl potential
field (PF)} where the $\alpha$-parameter vanishes $(\alpha=0)$,
(ii) a {\sl linear force-free field (LFFF)} $(\alpha = const)$, and
(iii) a {\sl nonlinear force-free field (NLFFF)} with a spatially
varying $\alpha({\bf r}) \neq 0$.

Previously we developed a nonlinear force-free field code that
models the 3-D coronal magnetic field in (flaring or non-flaring)
active regions, which we call the {\sl Vertical-Current Approximation
(VCA-NLFFF) code} (Aschwanden and Sandman 2010; Sandman et al.~2009;
Sandman and Aschwanden 2011;
Aschwanden et al.~2012, 2014a, 2014b, 2015, 2016c, 2018
Aschwanden 2013a, 2013b, 2013c, 2015, 2016, 2019b, 2019c;
Aschwanden and Malanushenko 2013; Warren et al.~2018).
It has an advantage over other NLFFF codes by
including the magnetic field of (automatically traced)
coronal loops, while standard NLFFF codes use the transverse
field components in the non-forcefree photosphere, which is
inadequate to reconstruct the force-free coronal magnetic field
(DeRosa et al.~2009). The inclusion
of coronal field directions is accomplished by automated tracing
of 2D-projected loop structures in EUV images, such as from AIA/SDO.

An approximate solution of the divergence-free and force-free
magnetic field can be derived analytically. The third version of 
the (VCA3-NLFFF) code is described in more detail in Aschwanden (2019c), 
while the latest version of this forward-fitting (VCA4-NLFFF) code
is described in more detail in Appendix A. 

The analytical
solution of the force-free magnetic field of a buried unipolar
magnetic charge is (Aschwanden 2019c), 
\begin{equation}
        B_r(r, \theta) = B_0 \left({d^2 \over r^2}\right)
        {1 \over (1 + b^2 r^2 \sin^2{\theta})} \ ,
\end{equation}
\begin{equation}
        B_\varphi(r, \theta) =
        B_0 \left({d^2 \over r^2}\right)
        {b r \sin{\theta} \over (1 + b^2 r^2 \sin^2{\theta})} \ ,
\end{equation}
\begin{equation}
        B_\theta(r, \theta) =
        B_0 \left({d^2 \over r^2}\right)
        {b^2 r^2 \sin^3(\theta) \over
        (1 + b^2 r^2 \sin^2 \theta )} {1 \over \cos \theta}
        \ ,
\end{equation}
\begin{equation}
        \alpha(r, \theta) \approx {2 b \cos{\theta} \over
        (1 + b^2 r^2 \sin^2{\theta})}  \ .
\end{equation}
where the constant $b$ is defined in terms of the number
of full twisting turns $n_{twist}$ over the loop length $L$,
i.e., $b = 2 \pi n_{twist}/L$. The azimuthal field component
$B_{\varphi}(r, \theta)$ (Eq.~8) is most suitable to measure
torsional oscillations of a twisted magnetic field directly
(Eq.~4). 

The total non-potential magnetic field from an arbitrary number
of $m=1,...,n_m$ unipolar magnetic charges can be 
approximately obtained from the vector sum of all magnetic
components $m$,
\begin{equation}
        {\bf B}({\bf x}) = \sum_{m=1}^{N_{\rm m}} {\bf B}_m({\bf x}) \ ,
\end{equation}
where the vector components ${\bf B}_m=(B_{x,m}, B_{y,m}, B_{z,m})$
of the non-potential field of a magnetic charge $m$ have to be
transformed into the same coordinate system.
The resulting model can be parameterized with $5 n_{m}$ 
free parameters $(B_m, x_m, y_m, z_m, \alpha_m)$ per magnetic
charge for a non-potential field, or with $4 n_{m}$ free 
parameters for a potential field (with $\alpha_m = 0$).
This new analytical solution is 
of second-order accuracy in the divergence-freeness condition, 
and of third-order accuracy in the force-freeness condition	
(Aschwanden 2019c).

\section{	OBSERVATIONS AND DATA ANALYSIS		}

\subsection{	Data Sets				}

Our goal is to study the time evolution of the free energy $E_f(t)$
and related energy dissipation mechanisms during solar flares.
For this purpose we use the line-of-sight magnetograms $B_z(x,y)$
from the {\sl Helioseismic and Magnetic Imager (HMI)}
(Scherrer et al.~2012), and 
EUV images from the {\sl Atmospheric Imaging Assembly (AIA)}
(Lemen et al.~2012),
both instruments onboard the {\sl Solar Dynamics Observatory (SDO)}
(Pesnell et al.~2011). We analyze the same subset of 11 GOES X-class 
flares presented in previous papers (Aschwanden et al.~2014a; 
Aschwanden 2019c), which represents all X-class flares observed 
with the SDO during the first 3.5 years of the mission
(2010 June 1 to 2014 January 31). This selection of events has a
heliographic longitude range of $[-45^\circ, +45^\circ]$, for which
magnetic field modeling can be faciliated without too severe
foreshortening effects near the solar limb. We use the 45-s
line-of-sight magnetograms from HMI/SDO. We make use of all
coronal EUV channels of AIA/SDO (in the six wavelengths
94, 131, 171, 193, 211, 335 \ang ), which are sensitive to
strong iron lines (Fe VIII, IX, XII, XIV, XVI, XVIII, XXI, XXIV)
in the temperature range of $T \approx 0.6-16$ MK. The spatial
resolution is $\approx1.6"$ (0.6" pixels) for AIA, and
the pixel size of HMI is 0.5". The coronal magnetic field is
modeled by using the line-of-sight
magnetogram $B_z(x,y)$ from HMI and (automatically detected)
projected loop coordinates $[x(s), y(s)]$ in each EUV wavelength of AIA.
A full 3-D magnetic field model ${\bf B}(x,y,z)$ is computed
for each time interval and flare with a cadence of 3 minutes, 
where the total duration of a flare is defined
by the GOES flare start and end times, including a margin
of 30 minutes before and after each flare. The size of
the computation box amounts to an area with a width and length
of 0.35 solar radius in the plane-of-sky, and an altitude range
of $h=0.003-0.25$ solar radius. The total number of analyzed data
includes 300 HMI images and 3538 AIA images.

\subsection{	Loop Decimation 			}

The automated tracing of loop segments in EUV images from AIA
yields typically $\approx 500$ loop segments in 6 different
coronal wavelengths per time frame, which may contain a sizable 
number of inaccurately traced loops, resulting from confusion by
over- and under-lying structures along the line-of-sight, 
or by contamination from {\sl moss-like features}
(De Pontieu et al.~1999). It is therefore recommendable to
filter out such spurious loop structures that exhibit a large
persistent misalignment angle during subsequent iterations in our
forward-fitting VCA4-NLFFF procedure. We define a threshold 
for the number of loops used in the fitting of
the VCA4-NLFFF code, expressed by an elimination ratio 
$q_{elim}=n_{elim}/n_{det}$ of the number $n_{elim}$ of
eliminated loops as a fraction of all detected $n_{det}$ loops.
The selection of eliminated loops is made by sorting of 
their misalignment angles $\mu_3$ (between the observed loop
direction and the theoretical magnetic field line direction) 
at each iterative time step of our forward-fitting (VCA4-NLFFF)
code. At the first iteration step, the theoretical model is
given by the potential field, while the theoretical model
converges towards a divergence-free and force-free solution
in the later iteration steps. 
A high elimination rate (say $q_{elim}=0.9$)
yields a small number of fitted loops, which results into a 
smaller misalignment angle $\mu_3$, while a low elimination rate
(say $q_{elim}=0.6$) yields larger statistics, but also larger
misaligment angles. The choice of the loop elimination rate
is therefore a trade-off between high accuracy and statistical
robustness. Consequently, the variation of the loop elimination rate
yields also a measure of the uncertainty of the derived free 
energies $E_f(t)$. 
 
\subsection{	Example of an Analyzed Event		}

An example of an analyzed event is given in Fig.~2, where also
the control parameters of the VCA4-NLFFF algorithm are listed. 
First, a HMI magnetogram (blue images in Fig.~2) is decomposed 
into $n_m=30$ unipolar sources, which yield the photospheric
magnetic field values $B_m$, the coordinates of the unipolar
magnetic charges $(x_m, y_m, z_m)$ below the photosphere, and
the nonlinear $\alpha_m$-parameters for each of the
$m=1,...,30$ magnetic charges. In the second task, a total of
$n_{det}=558$ loop structures are detected (yellow curves
in Fig.~2), while for each loop
structure the projected coordinates $[x(s_i), y(s_i)]$ along
the curvi-linear loop geometry are measured and interpolated
in $n_s=5$ segments along the loop trajectory $s_i,\ i=1,...,n_s$. 
The line-of-sight coordinate $[z(s_i]$ is 
then estimated from 50 parabolically curved loop geometries 
(for each loop),
where the best-fitting geometry that has the smallest
misalignment angle $\mu_3$ is chosen for further iterations
of the VCA4-NLFFF fitting process. The iterative forward-fitting
varies both the $n_m=30$ $\alpha$-parameters and the
$n_{loop}=223$ loop geometries, until it converges to a
minimum misalignment angle $\mu_2$. The run shown in Fig.~2
reached a minimized misalignment angle of $\mu_2=9.6^\circ$
after $n_{iter}=15$ iteration steps. A set of magnetic field
lines is shown in Fig.~2 (red curves), where all field lines
intersect with the midpoints of the observed loop segments 
(yellow curves). The full procedure
is then repeated for every time step with a cadence of 3 minutes,
which yields the time evolution of the free energy $E_f(t)$
as defined in Eq.~(1). Moreover, we repeated each run four times
with different loop elimination factors, i.e., $q_{elim}=
0.6, 0.7, 0.8, 0.9$, in order to estimate uncertainties of
the free energy $E_f(t)$.  

\section{	RESULTS					}

The results of our analysis is the time evolution of the
free energy, $E_f(t)$, which is shown for 11 GOES X-class
flares in Figs.~3 and 4 (red diamonds) on a logarithmic scale,
and in Fig.~5 on a linear scale (diamonds), while the best-fit
parameters are listed in Table 1. For comparison we show also
the free energy $E_f^W(t)$ obtained from the {\sl Wiegelmann 
nonlinear force-free field (W-NLFFF)} code as calculated earlier 
(Figs.~6 and 7 in Aschwanden 2019c),
the GOES 1-8 \ang\ flux (dashed curves in Figs.~3 and 4),
and the GOES time derivative (hatched areas in Figs.~3 and 4.)
Note that we calculated VCA4-NLFFF solutions with a cadence of 
3 minutes, while the used Wiegelmann NLFFF solutions have a
cadence of 12 minutes.

\subsection{	Free Energy Oscillations			}

The most striking result of this study is the discovery of
oscillations in the free energy of each of the 11 analyzed
large (GOES X-class) flares, as it can be seen by eye in
Figs.~3, 4, and 5. This data set of 11 flares contains
between $n_p=1$ and $n_p=4$ pulses during the duration
of a flare (Table 1).

Fitting a sinusoidal function with a linear trend,
\begin{equation}	
	E_f(t)^{osc} = A_0 + A_1 (t - t_0) + A_2 \sin{ \left(
	2 \pi (t - t_0) \over P \right)} \ ,
\end{equation}
to the observed time profiles of the free energy (Eq.~4) we find
oscillations periods in the range of $P=(10.0, ... ,22.5)$ minutes,
amplitudes of $A_0=(0.34, ..., 2.29) \times 10^{30}$ erg,
linear gradients of $A_1=(-0.50, ..., 2.78) \times 10^{30}$ erg
hrs$^{-1}$, oscillation amplitudes of 
$|A_2|=(0.04, ..., 0.73)\times 10^{30}$ erg, and
modulation depths of $M=|A_2/A_0|=0.09, ..., 0.42$, 
as listed in Table 1. The time intervals during which we fit 
the sinusoidal function are chosen include at least the
flare duration (given by the NOAA flare start times $t_s$ and 
end times $t_e$, marked with vertical dashed lines in Figs.~3, 4,
and 5), and are extended by margins of up to $\pm 0.5$ hrs. 
A measure of the adequacy of fits is given by the 
cross-correlation coefficient
\begin{equation}
	CCC = { \sum_i E_f^{osc}(t_i) E_f^{obs}(t_i) \over 
		\sum_i E_f^{osc}(t_i) \times \sum_i E_f^{obs}(t_i) } \ ,
\end{equation}
between the observed $E_f^{obs}(t)$ and the best-fit oscillatory
function $E_f^{osc}(t)$ (Eq.~12), which all are found in the
range of $CCC=0.67-0.90$, with a mean value of $CCC=0.83\pm0.07$. 
Consequently, the oscillations are significant and highly correlated 
with all observed data.

As a test of the 11 $>X1.0$ class flare events analyzed here,
we investigate also a small flare event \#003, which is 
of GOES class M, shown in Fig.~4f. This event shows a poor 
cross-correlation coefficient ($CCC = 0.47$), probably affected
by higher noise in the free energy solutions than in the
case of X-class flares.

In order to separate valid coronal loops
from spurious detections we repeated the fits for four different
elimination thresholds of ($q_{elim}=0.6, 0.7, 0.8, 0.9$) and selected
those loop data sets that yield the highest cross-correlation
coefficient. We list in Table 1 also the median misalinment angles 
$\mu_2 =5.3^\circ-14.7^\circ$ of the best-fitting VCA4-NLFFF 
magnetic field models, which represents a goodness-of-fit criterion
of the theoretical magnetic field model (Section 2). 

\subsection{	Phase Speed 					}

Let us estimate the phase speed of a hypothetical wave that produces
torsional oscillations, as depicted in Fig.~7 (left).
The example of flare \#351 shown in Fig.~2 has a field-of-view
of $x_{fov}=0.35$ solar radius and the separation between the
western and eastern sunspots amounts to $2 R \approx 130$ Mm, where 
we approximate the length $L$ of the field line by a semi-circular geometry, 
$L=\pi R \approx 200$ Mm. The propagation distance $L$ corresponds 
to the length of the mean field line between the two footpoint nodes
of the dominant dipole configuration in each flare. We measured
these length scales and find a mean of $L = 135 \pm 35$ Mm (Table 1).
Based on the assumption of the oscillations in the free energy $E_f(t)$ 
or azimuthal field component $B_f(t)$ to be in the fundamental standing mode, 
i.e., with the wavelength equal to $2L$ and the wave period $P_{obs}$ 
being the time scale between two subsequent peaks, the phase speed 
$v_{phase}$ can be estimated as, 
\begin{equation}
	v_{ph} = {2 L \over P_{obs}} \approx 315 \pm 120  
	\quad ({\rm km}\ {\rm s}^{-1}) \ . 
\end{equation}
The values of the lengths $L$, the time periods $P_{obs}$, and
the phase speeds $v_{ph}$ are listed for the analyzed 11 flare events
in Table 1.

\subsection{	Evolutionary Trends of Free Energy		}

Besides the oscillatory motion of the free energy $E_f(t)$, we note
also a linear trend $A_1 \neq 0$ in all flares, 
\begin{equation}
	E_f(t)^{lin} = A_0 + A_1 (t - t_0) \ ,
\end{equation}
which is positive in 7 cases, and negative in 4 cases (Table 1).
The signs of the linear trends $A_1$ in the free energy
$E_{\varphi}(t)$ bear important diagnostics whether
the free energy is dissipated (if negative) or if new emerging
flux (if positive) is occurring. By multiplying the linear energy
trend with the flare duration $D$, we obtain the net increase or
decrease of free energy during the flares,
\begin{equation}
	\Delta E_f = A_1 \ D \ .
\end{equation}
Note that the time evolution of the free energy $E_f(t)$ in the
{\sl Wiegelmann (W-NLFFF)} code shows a significant decrease in 
8 flare events
(Figs.~3 and 4, blue curves), but stays constant in the other
3 investigated cases (flares \#148, 220, 384). None of 
the flares exhibits oscillations when using the Wiegelmann
(W-NLFFF) code, which is a result of the
pre-processing and smoothing technique (which suppresses torque;
Wiegelmann et al.~2006),
as well as due to the lower cadence of 12 minutes available here. 

The time evolution of the free energy $E_f(t)$ oscillations
reveal that oscillations are detected during the entire 
analyzed time window of $[t_s-0.5, t_e+0.5]$ hrs 
in two cases (\#147, \#148), which would be expected
in the case of an omni-present instrumental effect.
However, the beginning of the oscillation period is 
detected in 8 cases (events \#12, \#37, \#66, \#67, 
\#220, \#349, \#351, and \#384, see Fig.~5), which
clearly demonstrates the solar origin of the triggered
oscillations, rather than being an omni-present instrumental
effect. 

\subsection{	Comparison of NLFFF Codes 		}

The comparison with the Wiegelmann NLFFF code shows some
intriguing differences.  The absolute value of the 
free energy matches those of the Wiegelmann NLFFF code
only in 2 cases out of the 11 events (flares \#66 and \$67
in Figs.~3c and 3d), but is otherwise always lower, i.e.,
$E_f(t) \le E_f^W(t)$, up to an order of magnitude 
(e.g., flare \#147 in Fig.~3e). The reason for this discrepancy
is not understood at this time. The most likely reason is that
no coronal loops have been detected (or they have been eliminated)
in some compact high magnetic field regions, using our automated 
loop tracing code, especially
in the core of the active region, where ``moss strucutres'' 
interfere with the tracing of overlying loops. However, we have
to be aware that the free energy is a small difference of two
large quantities, so that the relative uncertainty of free
energy measurements is much larger than for the potential
or nonpotential field components. Typically, the free energy
amounts to $\approx 1\%$ to 10\% of the nonpotential energy,
and thus an order of magnitude difference in the free energy
corresponds only to 1\% to 10\% of the nonpotential energy.
In a previous study we found that both the potential energy
and nonpotential energy agree within $E_p/E_p^W=0.8\pm0.2$ 
and $E_{np}/E_{np}^W=0.8\pm0.2$ with the Wiegelmann NLFFF code
(see Fig.~6 in Aschwanden 2016).
Alternatively, the pre-processing of the magnetogram data
carried out with the Wiegelmann NLFFF code (Wiegelmann 
et al.~2006) could possibly over-estimate the free energy 
(Yan Xu, private communication).

\section{	INTERPRETATION AND DISCUSSION		}

\subsection{	Instrumental Oscillatory Effects	}

Since these observations report for the first time oscillations
in the free (magnetic) energy of solar flares, we discuss 
first possible effects of instrumental origin. The range of
observed oscillations (in 11 flare events) cover the range of
$P \approx 10-22$ min and have a mean and standard deviation of
\begin{equation}
	P_{obs}= 15.1 \pm 3.9 \ ({\rm min}) \ .
\end{equation}

Since the SDO spacecraft is positioned in the Lagrangian point L1,
there are no eclipsing time intervals of the spacecraft orbit
that could modulate the observed flux.

Since the Sun is rotating, any tracking of a particular heliographic
position of a solar flare or an active region needs to be interpolated
between two pixels, which could possibly introduce a pixel-step modulation.
We estimate the time scale of a pixel shift based on the synodic solar 
rotation rate,
\begin{equation}
	\Delta t_{pix} = P_{syn} \left( {\Delta x_{pix} \over 2 \pi R_\odot} \right) 
	\approx 3.2\ ({\rm min}) \ ,
\end{equation} 
which amounts for a synodic rotation time of $P_{syn}=27.27$ days, an HMI pixel
size of $\Delta x_{pix}=0.5\arcsec$, and an average solar radius of 
$R_{\odot} \approx 960\arcsec$, to a time interval of $\Delta t_{pix}=3.2$ min
for HMI, or $\Delta t_{pix}=3.9$ min for AIA pixels. The beat frequency between 
the two instruments amounts to a time scale of 
\begin{equation}
	P_{beat}  = \left[ {1 \over P_{HMI}} - {1 \over P_{AIA}} \right]^{-1} 
		  = \left[ {1 \over 3.2 {\rm (min)}} - {1 \over 3.9 {\rm (min)}} \right]^{-1}
		  = 17.8 \ {\rm (min)} \ .
\end{equation}
Although the beat frequency is close to the frequency of the observed mean
oscillation period (Eq.~17), 
sub-pixel variations of the sensitivity (or modulation
transfer function) of AIA are unlikely to explain the observed oscillations, 
since the width of the point spread function is substantially wider 
($1.5\arcsec$) than the pixel size ($0.6\arcsec$) 
(Lemen et al.~2012; Boerner et al.~2012; Grigis et al.~2012). 
The point-spread function may also imply a slight sub-pixel variation of the
sensitivity of CCD pixels (between the pixel center and pixel edge). 
We calculated this variation by pixel-wise superposition of point-spread 
functions and found an inter-pixel variation of $\lapprox$3\%. Moreover, the
rms variation in the AIA flat-field is found to be of order 2\% 
(Boerner et al.~2012).
The point-spread function was also measured during a lunar transit,
finding that the PSF of AIA/SDO is better by a factor of two than
the EUV telescope onboard of TRACE (Poduval et al.~2013).
In summary, no instrumental effect is known from AIA or HMI that could 
modulate the observed free energy evolution $E_f(t)$ with periods of 
$P \approx 10-20$ min (James Lemen, Paul Boerner, Mark Cheung, Wei Liu; 
private communication). 

Incidentally we did choose a cadence of 3 min for the AIA images
with the VCA4-NLFFF code, while a cadence of 12 min is available from the
W-NLFFF code using HMI data (Aschwanden 2019c). The cadence of 12 min is 
close to the mean observed period of $P_{obs}=15.1 \pm 3.9$ min, and thus 
the observed oscillations are fully resolved with the 3-minute cadence of 
the VCA4-NLFFF code only, while they are not resolved with the available data 
from the W-NLFFF code (Figs.~3 and 4).

\subsection{	Torsional Oscillations of Alfv\'enic MHD Waves  }

After we ruled out instrumental effects, we turn to a physical model
for the interpretation of the observed oscillations. Since the free
energy is related to the azimuthal magnetic field component by
definition, $B_f(t)=B_{\varphi}(t)$, in our vertical-current
approximation model (i.e., VCA4-NLFFF code), we can interpret
the observed pulsations in terms of {\sl torsional oscillations}
of a helically twisted magnetic field structure in the flaring 
active region. The overall magnetic field structure in an active 
region can often be approximated with a bipolar structure 
in approximate East-West direction, modeled with two unipolar
magnetic charges of opposite magnetic polarity. If this bipolar
structure is near to a potential field configuration, the magnetic
field lines connect the leading sunspot with the trailing region
of opposite magnetic polarity along the most direct dipolar field lines,
while these field lines become helically twisted like a sigmoid
in strongly nonpotential field configurations (Aschwanden 2004, 2019a). 
In reality, the magnetic field
structure consists of many more smaller dipoles and quadrupoles,
but the overall bipolar structure is dominant. A good example is
shown in Fig.~2, showing a major dipolar configuration between
the western positive magnetic polarity (leading sunspot) and 
the eastern negative magentic polarity (trailing plage). 

For the interpretation of the wave type we may consider
the MHD modes predicted in a standard thin fluxtube model,
which includes slow-mode, fast-mode MHD waves, and torsional 
Alfv\'en waves (Edwin and Roberts 1983; Robert et al.~1984).
Our estimate of the phase speed
yields a mean value of $v_{phase}=315 \pm 120$ km s$^{-1}$
(Section 4.2 and Table 1), which is inbetween the phase speed
expected for slow-mode MHD waves, defined by the sound speed
$c_s = 150 \sqrt{ {T_e / 1\ {\rm MK)}} }$ (km s$^{-1}$),
and the torsional Alfv\'en waves, defined by the Alfv\'en speed
$v_A = 2.18 \times 10^6 B/\sqrt{n_e}$ (km s$^{-1}$). Typical values
for the Alfv\'en speed in active regions are of order $v_A\approx
1000$ km s$^{-1}$ (using $B \approx 15$ G and $n_e \approx 10^9$ cm$^{-3}$),
while the Alfv\'en speeds in flaring loops have a lower value of
$v_A \approx 300$ km s$^{-1}$, for instance assuming an order of 
magnitude higher densities than the surrounding backgroun corona 
($n_e \approx 10^{10}$ cm$^{-3}$).  
Therefore, our measured phase speed, $v_{phase}=315 \pm 120$ km s$^{-1}$
is consistent with torsional Alfv\'en waves in flare loops with enhanced
electron densities. On the other side, (fast) kink-mode periods have 
been found to have shorter periods ($P_{kink} = 5.4 \pm 2.3$ min; 
Aschwanden et al.~2002), while slow-mode periods were found to have 
similar periods, ($P_{slow} = 17.6 \pm 5.4$ min; Wang et al.~2003), 
than measured here, but oscillation periods depend on both the
loop lengths $L$ an the phase 
speed $v_{phase}$ of the MHD waves ($P \approx 2 L / v_{phase}$),
and thus do not provide an independent diagnostic of the wave type. 

Alternatively, the phase speed would be
consistent with slow-mode MHD waves also, if the temperature in
the oscillating loops amounts to $T_e=4$ MK, yielding
$v_{phase} \approx c_s \approx 300$ km s$^{-1}$, but such an 
interpretation would face a number of inconsistencies:
i) Theoretically, the coupling between torsional Alfv\'enic
oscillations and slow-mode waves is very weak in the low 
plasma-$\beta$ regions of the solar corona (De Moortel et al.~2004;
Zaqarashvili et al.~2006; Afanashvili et al.~2015); 
(ii) The restoring force in slow-mode MHD oscillations
is mainly the plasma pressure, rather than the Lorentz force in the case
of torsional Alfv\'enic oscillations. If slow-mode oscillations
are the case here, some correlation between the radiated
soft X-ray flux and free energy oscillations would be expected,
which is not obvious in the GOES data and free energy time profiles
shown in Figs.~3 and 4; and 
(iii) The damping of slow-mode oscillations has been found to
be strong (Wang et al.~2003a, 2003b) due to the conductive 
losses from hot postflare loops, opposed to the mostly
decay-less amplitudes observed here (Fig.~5).

A further test of the hypothesis of torsional waves is
the self-consistency of the time evolution of the various
magnetic energies. We show the time evolution of the
potential energy $E_{p}(t)$ and the free energy $E_f(t)$
in Fig.~6, while the evolution of the free energy $E_f(t)$
is shown in Fig.~5. The potential energy is uncorrelated 
to the free energy, as the cross-correlation coefficients 
in Fig.~6 demonstrate, with a mean and standard deviation 
of $CCC = 0.26 \pm 0.16$. This uncorrelatedness is expected,
because the azimuthal magnetic field component is
orthogonal to the radial magnetic potential field component,
i.e., $B_{\varphi}(t) \perp B_r$ (Fig.~7 left). In contrast, 
if the potential energy and free energy would be correlated,
the azimuthal energy would vary proportionally with the
potential field energy, which could only be explained by
field-aligned current oscillations (Fig.~7 right).
At the same time, the uncorrelatedness of the potential
energy and free energy rules out instrumental effects
as the source of the observed oscillations. 

Theoretically, the dispersion
relation and coupling of wave modes in twisted magnetic 
flux tubes is treated in Chapter 7 of Roberts (2019).  
The process of resonant energy transfer from torsional
oscillations into acoustic waves has been modeled also 
(Zaqarashvili 2003; Zaqarashvili and Murawski 2007).
Further analytical models and numerical simulations of
torsional waves in flux tubes and the generation of
compressible flows have been carried out in
a number of studies (e.g., Copil et al.~2008, 2010;
Vasheghani-Farahani et al. 2010, 2011), which may
explain swirling motions, tornadoes, and helical
spirals in twisted solar structures 
(Mozafari Ghoraba and Vasheghani-Farahani 2018; 
Vasheghani-Farahani et al.~2017). 

\subsection{	The Lorentz Force 			    } 

Torsional oscillation in magnetic flux tubes are usually
referred to the oscillations with the Lorentz force as a
restoring force. In the observations presented here, the 
oscillation of the azimuthal magnetic field component
$\Delta B_{\varphi}^2(t) \propto \Delta E_f(t) \propto \sin{(t)}$ 
appears to cause 
small-amplitude torsional oscillations with a Lorentz force acting
as a counter balance in restoring the toroidal motion.
Since this torsional system exhibits a non-constant motion, the 
Lorentz force cannot be zero, and thus the magnetic system
cannot be force-free. This appears to be in contrast to the
near-forcefree solutions attempted with the vertical-current 
approximation
(VCA4-NLFFF) code. This code is designed to converge to a
divergence-free and force-free solution for every time step.
Obviously our solutions of the VCA4-NLFFF code are not 
exactly force-free, but oscillate in synchrony with the
Lorentz force driver. In fact, the time evolution shown 
in Fig.~6 explicitly demonstrates that the magnetic potential 
field is almost constant and thus force-free, while the
free energy oscillates and thus is non-forcefree. 
The degree of non-force-freeness is quantified with the
2-D and 3-D misalignment angles (typically 
$\mu_2 \approx 10^\circ-20^\circ$ in our study).
One would expect that an equilibrium situation would occur when the
small-amplitude torsional oscillations decay to a minimum
energy state, with both the potential field and the
free energy becoming constant, while the Lorentz forces
converge to zero.

\subsection{Measuring Torsional Oscillations in the Corona}

Torsional oscillations of sunspots, which harbour the
footpoints of many active region loops have been observed to
rotate on much longer time scales. For instance,
rotational periods in the range of $P \approx 2-8$ days 
were reported (Gopasyuk and Kosovichev 2011; 
Grinon-Marin et al.~2017), and thus cannot be 
considered as the driver of the azimuthal oscillations 
observed here (with $P\approx 10-22$ min), although 
they are located at the right place. 

The torsional motion of coronal loops is difficult to
measure, because we can observe their 2-D projected 
motion only, unless we attempt a 3-D reconstruction 
by stereoscopic means or by 3-D magnetic field modeling. 
Alternatively, it was proposed to detect periodical
broadenings of spectral lines that result from periodic
azimuthal velocities (Zaqarashvili 2003; 
Zaqarashvili and Murawski 2007).
Ubiquitous torsional motions in type II spicules were
identified from spectral line velocities (of order
25-30 km s$^{-1}$), in particular in the outer red 
and blue wings of chromospheric lines, besides the
field-aligned flows ($\approx 50-100$ km s$^{-1}$)
and swaying motions ($15-20$ km s$^{-1}$)
(De Pontieu et al.~2012; Sekse et al.~2013).
However, these velocities are generally smaller 
than those measured here.

\subsection{	Damping of Torsional Waves	}

In the time evolution of oscillatory motion shown in Fig.~5
we note that most oscillation events do not
exhibit a detectable degree of wave damping, which 
creates a new puzzle.
Most MHD oscillation modes exhibit relatively short 
damping times, in the order of 1-4 pulses 
(Aschwanden et al.~1999; Nakariakov et al. 1999),
while we find $n_p=1-4$ pulses per flare duration here 
without damping (Table 1).
One damping mechanism is resonant absorption, which
transfers energy from the fast kink mode to Alfv\'enic
azimuthal oscillations within the inhomogeneous layer
(Ruderman and Roberts 2002). However, there also cases
with no detectable damping in the post-flare phase,
called {\sl decay-less} oscillations (Anfinogentov et al.~2013;
Nistico et al.~2013).
The damping due to resonant absorption (acting in the inhomogeneous
regions of a flux tube where energy is transferred from the kink mode
to Alfv\'en azimuthal oscillations) is analytically
treated in Ruderman and Robers (2002), who suggest that those loops
with density inhomogeneities on a small scale (compared with the
loop cross-sectional width) are able to support (observable) coherent
oscillations for any length of time, while loops with a smooth cross-sectional
density variation do not exhibit pronounced oscillations.
An alternative explanation is that torsional waves efficiently 
modulate gyrosynchrotron emission,
which are not subject to radiative and conductive damping and this way can
produce stable pulse repetition rates in solar radio pulsation
events during flares (Tapping 1983). 

\section{	CONCLUSIONS				}

Torsional oscillations, which essentially are dynamical systems
that have the torque as restoring force, have been invoked in many
solar phenomena, such as for the global Hale 11-year solar cycle, 
oscillatory deviations from the solar differential rotation, 
meridional circulation, the solar and stellar convection zone,
helioseismic measurements of flows in the solar interior,
photospheric flux tubes, flows of intergranulary magnetic 
flux concentrations, sunspot rotations, torsional motions
of coronal loops and threads, type II spicules, solar tornadoes,
solar radio bursts, and torsional Alfv\'en waves in the solar wind. 
Although detailed models of these solar phenomena require
different physical mechanisms, the fundamental concept of
circular motion with interacting torque is common.

Here we adding a new discovery (to our best knowledge) that
exhibits torsional motion in its purest form, because the
oscillatory azimuthal rotation can directly be measured
from a 3-D magnetic field model in terms of the azimuthal 
field component $B_\varphi(t)$. Moreover, the 3-D magnetic
field model has a high degree of observational fidelity,
because the divergence-free and force-free magnetic field
solution is fitted to the observed coronal loops, which
supposedly represent the ``true'' magnetic field 
in a low plasma-$\beta$ corona. Traditional NLFFF models
extrapolate the coronal field from a non-forcefree
photospheric field, after pre-processing of the photospheric
boundary. Moreover, the pre-processing is designed to 
minimize forces and torques, and this way smoothes out any
torsional signal. Consequently, torsional oscillations cannot
be reconstructed with pre-processed data (besides the
insufficient temporal cadence of 12 min used here),
and thus have never been detected with traditional
NLFFF codes.

Torsional oscillations have been identified here for a
complete data set of all investigated 11 (GOES $\ge$X1.0-class) 
flares, without any selection effects. 
The identification of torsional oscillations is
based on relatively high cross-correlation coefficients
($CCC \approx 0.83\pm0.07$) between the observed time evolution
of the free energy $E_f(t)=E_\varphi(t)$ with an oscillatory
sine function, and the agreement of the mean observed phase speed, 
$v_{phase}=315 \pm 120$ km s$^{-1}$, with torsional 
Alfv\'enic MHD waves in density-enhanced flare loops. 
Our interpretation suggests that the torsional wave may 
couple with the fast-mode MHD wave, rather than with the
slow-mode MHD wave. Also, the torsional oscillations 
are found to be decay-less, similar to some fast kink-mode
decay-less events.

Besides the oscillatory motion, the time evolution of the
free energy $E_f(t)$ shows also a linear trend, which 
increases or decreases the free energy in a linear fashion.
The simple concept that the free energy drops like a
step function during the energy dissipation phase of a
flare appears to be over-simplified, because we detect
multiple pulses, each one accompanied by a rise and 
decay of the free energy (Aschwanden 2019c). Consequently,
we can filter out the oscillatory part and determine
the free energy before (at time $t_s$) and after a flare
(at time $t_e$) from the linear trend, i.e., 
$\Delta E_f = E_f(t_s) - E_f(t_e)$, where flare energy dissipation
or magnetic flux cancelling occurs when $\Delta E_f < 0$ 
is negative, while emergence of new magnetic flux is 
required when $\Delta E_f > 0$.

Future studies should investigate whether torsional
oscillations occur during large flares only, what
the damping mechanism is for decay-less oscillation
events, how accurately we can measure and predict
the distributions of sound speeds in a flaring
active region, how torsional oscillations couple
with slow-mode MHD waves, and how torsional oscillation 
velocities agree with other methods, such as 
spectral line broadening. 

\acknowledgements
The author thanks for helpful discussions with 
Paul Boerner, Jim Lemen, Bart De Pontieu, Mark Cheung, 
Marc DeRosa, Wei Liu, and John Serafin.
Part of the work was supported by
NASA contract NNG 04EA00C of the SDO/AIA instrument and
the NASA STEREO mission under NRL contract N00173-02-C-2035.
The work of TW was supported by NASA grants 80NSSC18K1131 
and 80NSSC18K0668, and the NASA Cooperative Agreement 
NNG11PL10A to CUA.

\clearpage
\section*{APPENDIX A: The Vertical-Current Approximation Code 
		Version 4 (VCA4-NLFFF)	}

The {\sl Vertical-Current Approximation (Version 4) Nonlinear
Force-Free Field (VCA4-NLFFF)} code has been improved in a
number or ways over the last 6 years, starting from the original definition 
(Aschwanden 2013a; 2013b; Aschwanden and Malanushenko 2013c),
followed by more detailed code descriptions and performance tests
(Aschwanden 2016), more accurate analytical solutions 
in Version 3 (Aschwanden 2019c), and improved iteration schemes
in the current Version 4, which are described in this Appendix. 
 
The VCA4-NLFFF code is publicly available in the 
{\sl Solar SoftWare (SSW)} package $\$/$SSW/package
/mjastereo/idl/ and consists of a sequence of 13 modules, 
written in {\sl Interactive Data Language (IDL)}, which we describe 
here in turn, while a tutorial is provided on the webpage 
{\sl http://www.lmsal.com/ 
$\sim$aschwand /software/nlfff/nlfff$\_$tutorial.pro}.

{\underbar{(1) NLFFF$\_$CAT}} generates an event catalog in form of
a time series (defined by the starting time, total duration, and cadence 
of an event) with heliographic positions that follow the solar rotation. 
The longitude change $dl/dt$ per time cadence $P_{cad}$ is calculated 
from the mean synodic rotation rate $P_{syn}=27.2753 \times 86400$ s,
$$
	{dl\over dt} = 360^\circ \ \left( {P_{cad} \over P_{syn}} \right) \ .
	\eqno(A1)
$$

{\underbar{(2) NLFFF$\_$INIT}} reads the times, wavelengths, and 
heliographic positions of an event from the catalog file, 
saves control parameters, and calculates 
the cartesian coordinates (in units of solar radii from Sun center) 
for each time step and (quadratic) field-of-view (FOV), centered at 
the initial heliographic location and following the differential
solar rotation. The cartesian coordinate sytem $(x,y,z)$
is centered at Sun center and takes the curvature of the solar surface
or photospheric magnetogram automatically into account, with the
line-of-sight coordinate $z_{phot}$ of the photosphere defined by
$$
	z_{phot}(x,y) = \left( r_{phot}^2-x^2-y^2 \right)^{1/2} \ ,
	\eqno(A2)
$$
and the normalization given by the solar radius,
$r_{phot}=R_{\odot}=1$.

{\underbar{(3) NLFFF$\_$HMI}} reads the HMI/SDO magnetogram
$B_z(x,y)$ and writes an uncompressed FITS file. Note that a
scaling factor BSCALE=0.1 has been introduced in the FITS
header in recent years, which changes the absolute scaling of 
the magnetic field by an order of magnitude.

{\underbar{(4) NLFFF$\_$AIA}} reads AIA/SDO images 
in up to 8 wavelengths within a time interval of 12 s
for every time step (defined by starting time and cadence),
and writes uncompressed FITS files for each wavelength.

{\underbar{(5) NLFFF$\_$FOV}} displays the selected
FOV (at coordinates $[x_1,y_1,x_2,y_2]$ for the chosen
``cut-out'' subimage)
and caclulates the solar radius and the AIA pixel size
from solar ephemerids, which 
varies by $\approx 10\%$ due to solar distance variation
of the Lagrangian point (L1), where the SDO spacecraft is
located. 

{\underbar{(6) NLFFF$\_$MAGN}} extracts the line-of-sight
magnetogram $B_z(x,y)$ within the selected FOV coordinates,
and coaligns the FOV image (by shift and rotation) according 
to the pointing information given in the FITS header. 
The extracted HMI magnetogram is rebinned to $3 \times 3$
HMI pixels, and then a sequential decomposition of the
HMI magnetogram into unipolar magnetic charges is carried
out. The decomposition yields the 3-D coordinates 
$(x_i, y_i, z_i), i=1,...,n_m$ of sub-photospheric 
magnetic charges that are obtained by fitting the local 
potential field of a unipolar magnetic charge (which 
decreases with the squared distance, $B(r) \propto r^{-2}$).
The iterative algorithm, which starts with the largest
sunspot in the FOV and fits progressively smaller
magnetic field concentrations, is decribed in more detail
in Aschwanden and Sandman 2010), Aschwanden et al.~(2012).
Correction factors for re-binning and magnetic field
energies are computed also. 

{\underbar{(7) NLFFF$\_$TRACING}} reads AIA images and
performs automated tracing of curvi-linear structres, mostly
rendering coronal loop segments in the lowest hydrostatic
scale height of the corona. The algorithm starts with the
brightest structure in an AIA image and traces the local
ridge at position $[x(s_i), y(s_i)], i=1,...,n_s$ 
by linear and constant-curvature extrapolation 
from loop coordinate $s_i$ to $s_{i+1}$, where $n_s$ is the
number of points along a loop, calculated with a step $\Delta s$
corresponding to the pixel size. Some control parameters
of this algorithm include the low-pass filter $n_{sm1}$ (which
enhances faint loops relative to the local background), the
high-pass filter $n_{sm2}=n_{sm1}+2$, which eliminates broad
background structures, the minimum loop length
$l_{min}$, the minimum curvature radius $r_{min}$, a
threshold $q_{thresh1}$ of the intensity image $I(x,t)$, as well as a
threshold $q_{thresh2}$ of the filtered image $I(x,t)-smooth[I(x,t)]$,
a tolerance gap $n_{gap}$ in tracing along a loop ridge, etc.
More detailed descriptions of the automated loop tracing algorithm 
(also called {\sl Oriented Coronal Curved Loop Tracing (OCCULT)}
are given in Aschwanden et al.~(2008; 2013) and Aschwanden (2010). 

{\underbar{(8) NLFFF$\_$FIT}} reads the 2-D loop coordinates $[x(s_i),
y(s_i)]$ for each of the multiple-wavelength AIA images, which typically
encompasses $n_{loop} \approx 500$ loops. For each loop we
test a proxy of the 3-D coordinate $[z(s_i)]$, which has the geometry 
of a parabolic loop segment,  
$$
	h_{ijk}(s)= qh_{ijk} - (4\ qh_{ijk}) (s-0.5)^2 \ ,
	\eqno(A3)
$$ 
where $s$ is the normalized loop coordinate $0 < s_i < 1, i=0,...,n_s$, 
$qh$ is the normalized height coordinate $0 < h_j < 1, j=0,...,n_h$, 
while $s_1=0$ is the first loop footpoint,
$s_2=s_i$ is the second loop footpoint,
and $k=0,1$ is the order of the two conjugate footpoints.
With the third coordinate $z(s)$,
$$
	z(s) = \left[ (1 + h_{ijk}(s))^2 - x(s)^2 - y(s)^2 \right]^{1/2} \ ,
	\eqno(A4)
$$
we have a full 3-D model $[x(s), y(s), z(s)]$
of a traced loop segment, which converges by selecting those 
loop geometries $(i,j,k)$ that produce the smallest misalignment
angles between the 3-D magnetic field line vectors (Eq.~11) and the
observed 3-D loop directions.

An example of these trial loop geometries is given in Fig.~8,
Typically, we used $n_s=5$ segments per loop and
$n_h=5$ heights in the range of $h=0.003,...0.25\ R_{\odot}$,
which implies $n_{geo}=2 n_s n_h=50$ trial geometries. 
This proxy assumes that at least one loop footpoint is located
near the chromosphere, while the automated tracing detects an
arbitrary part of the remaining loop segment.
The additional constraint of limiting the height range to the
2-D projected loop length yields improved results, i.e.,
$|s_{n_s}-s_1| < (h_2-h_1)$. 

The code optimizes three tasks of the full 3-D model of
the magnetic field simultaneously: (i) the height model $h(s)$
for each loop, (ii) the 
nonlinear force-free field parameters $\alpha_m, m=1,...,n_m$
of sub-photospheric magnetic charges that represent
the nonpotential field, and (iii) the loop decimation as
described in Section 3.2. These three tasks contain a number
of recent optimizations that improve the convergence of the VCA4-NLFFF 
code and are superior to the previous VCA3-NLFFF code.
A more stable convergence (based on comparing the solutions in 
subsequent time steps) was also found by increasing the number
of fitted unipolar magnetic charges progressively with the
number of iterations. This way, the strongest sunspots dominate
the solutions. The optimization of $\alpha$-parameters is done 
by second-order extrapolation, while it was done by first-order
extrapolation in the previous VCA3-NLFFF code.
Typically, only 10-20 iteration steps are needed for 
final convergence, which is measured by the misalignment
angle in 3-D ($\mu_3$) or 2-D ($\mu_2$).
The optimization of the entire 3-D model is characterized
by a single parameter, the median value of the 3-D misalignment 
angle ($\mu_3$): 
$$
	\mu_3({\bf x}) = cos^{-1}
	\left(  {\bf B}^{model}({\rm x}) \cdot 
	        {\bf B}^{obs}({\rm x})  
	\over 
	        |{\bf B}^{model}({\rm x})| \cdot 
	        |{\bf B}^{obs}({\rm x})  | \right) \ .  
	\eqno(A5)
$$

{\underbar{(9) NLFFF$\_$FIELD}} calculates the magnetic field
lines of the nonlinear force-free
field solution, which is fully constrained by the $5 n_m$
coefficiencts $(B_m, x_m, z_m, \alpha_j), m=1,...,n_m$,
which typically amounts with $n_m=30$ to $5 n_m = 150$ coefficients.
We display a graphical rendering of the magnetic field lines by
showing magnetic field lines that intersect the midpoint of the
observed loops, which allows for a direct inspection of the
misalignment angles $\mu_2$.

{\underbar{(10) NLFFF$\_$CUBE}} calculates the nonlinear force-free
field $[B_x(x,y,z), B_y(x,y,z), B_z(x,y,z)]$ 
in a 3-D computation box $(N_x \times N_y \times N_z)$ that
is given by the photospheric FOV area
and a height range of $\Delta h=h_{max}-h_{min}=0.250-0.003$
solar radius. However, it is much more economical to store the
$5 n_m$ coefficients of our model rather than a full 3-D cube
with one-pixel resolution.

{\underbar{(11) NLFFF$\_$MERIT}} calculates figures of merit of
a VCA4-NLFFF solution, including measures of the divergence-freeness,
force-freeness, and weighted current.

{\underbar{(12) NLFFF$\_$ENERGY}} calculates the 
potential energy $E_p=\int [B_p(x,y,z)^2 / 8\pi] dV$, 
the nonpotential energy $E_{np}=\int [B_{np}(x,y,z)^2 / 8\pi] dV$, 
and the free energy $E_f=E_{np}-E_p$. Correction factors due to
rebinning and magnetic modeling are included.

{\underbar{(13) NLFFF$\_$DISP}} provides renderings of the
observed 2-D loops, reconstructed 3-D loops, best-fit field lines
of the VCA4-NLFFF solutions, viewed from two orthogonal directions,
and histograms of the 2-D and 3-D misalignment angles
(for instance see Fig.~2).

\clearpage

%%%%%%%%%%%%%%%% REFE %%%%%%%%%%%%%%%%%%%%%%%%%%

\section*{	References	}	%REFERENCED

\def\ref#1{\par\noindent\hangindent1cm {#1}}

\ref{Afanasyev, A.N. and Nakariakov, V.M. 2015,
	AA 582, A57.}
\ref{Anfinogentov, S., Nistico, G., and Nakariakov, V.M. 2013,
        AA 560, 107.}
\ref{Antolin, P. and Shibata, K. 2010, ApJ 712, 494.}
\ref{Aschwanden, M.J., Fletcher, L., Schrijver, C., and Alexander, D.
 	1999, ApJ 520, 880.}
\ref{Aschwanden, M.J., DePontieu, B., Schrijver, C.J., and Title, A.
 	2002, SoPh 206, 99.}
\ref{Aschwanden, M.J.,
        2004, {\sl Physics of the Solar Corona. An Introduction},
        Praxis and Springer, Berlin, 216.}
\ref{Aschwanden, M.J. and Sandman, A.W.
        2010, AJ 140, 723} % (VCA)
\ref{Aschwanden, M.J. 2010, SoPh 262, 399.}
\ref{Aschwanden, M.J., Wuelser, J.P., Nitta, N.V., Lemen, J.R., 
	Schrijver, C.J., et al. 2012, ApJ 756, 124} % (VCA)
\ref{Aschwanden, M.J.
        2013a, SoPh 287, 323} %(VCA I)
\ref{Aschwanden, M.J.
        2013b, SoPh 287, 369} %(VCA III)
\ref{Aschwanden, M.J.
        2013c, ApJ 763, 115}  %(VCA)
\ref{Aschwanden, M.J., De Pontieu, B., and Katrukha, A. 2013,
	Entropy 15, 3007.}
\ref{Aschwanden, M.J. and Malanushenko, A.
        2013, SoPh 287, 345} %(VCA II)
\ref{Aschwanden, M.J., Xu, Y., and Jing J.
        2014a, ApJ 797:50.} %(Paper I)
\ref{Aschwanden, M.J., Sun, X.D., and Liu,Y.
        2014b, ApJ 785, 34} %(VCA)
\ref{Aschwanden, M.J.
        2015, ApJ 804, L20} %(VCA Magnetic Dissipation during 2019 March 29 flare)
\ref{Aschwanden, M.J., Schrijver, C.J., and Malanushenko, A.
        2015, SoPh 290, 2765} % (VCA) Blind stereoscopy
\ref{Aschwanden, M.J.
        2016, ApJSS 224, 25} %(VCA)
\ref{Aschwanden, M.J., Reardon, K., and Jess, D.
        2016c, ApJ 826, 61} %(VCA)
\ref{Aschwanden, M.J., Gosic, M., Hurlburt, N.E., and Scullion, E.
        2018, ApJ 866, 72} %(VCA)
\ref{Aschwanden, M.J.
        2019a, {\sl New Millennium Solar Physics}, Springer Nature,
        Switzerland, Science Library Vol. 458}
\ref{Aschwanden, M.J.
        2019b, ApJ 874, 131} %Helical twisting number and braiding 
\ref{Aschwanden, M.J. 2019c, ApJ 885:49.}
\ref{Boerner, P., Edwards, C., Lemen, J., Rausch, A.,
	Schrijver, C., Shine, R., Shing, L., Stern, R., et al.
	2012, SoPh 275:41.}
\ref{Chakraborty, S., Choudhuri, S., and Chatterjee, P. 2009,
	Phys.Rev.Lett. 102, 041102.}
\ref{Copil, P., Voitenko, Y. and Goossens, M. 2008, AA 478, 921.}
\ref{Copil, P., Voitenko, Y. and Goossens, M. 2010, AA 510, A17.}
\ref{Covas, E., Takakov, R., Moss, D., and Tworkowski, A. 
	2000 AA 360, L21.}
\ref{Covas, E., Moss, D., and Tavakol, R. 2004, AA 416, 775.}
\ref{De Moortel, I., Hood, A.W., Gerrard, C.L., and Brooks, S.J.
	2004, AA 425, 741.}
\ref{De Pontieu, B., Berger, T.E., Schrijver, C.J., and Title, A.M.
 	1999, SoPh 190, 419.}
\ref{De Pontieu, B., Carlsson, M., Rouppe van der Voort, L.H.M.,
	Rutten, R.J., Hansteen, V.M., and Watanabe, H.
	2012, ApJL 752, L12.}
\ref{DeRosa, M.L., Schrijver, C.J., Barnes, G., Leka,K.D., 
	Lites, B.W., Aschwanden, M.J., Amari, T., Canou,A., et al.
	2009, ApJ 696, 1780.} 
\ref{Durney, B.R. 2000, SoPh 196, 1.}
\ref{Edwin, P.M. and Roberts, B. 1983, SoPh 88, 179.}
\ref{Fedun, V., Verth, G., Jess, D.B., and Erdelyi, R. 2011,
	ApJL 740, L46.}
\ref{Gonzalez-Hernandez, I., Howe, R., Komm, R., and Hill, F.,
	2010, ApJL 713, L16.}
\ref{Gopasyuk, S.I. and Kosovichev, A.G. 2011, ApJ 729, 95.}
\ref{Gosling, J.T., Teh, W.L., and Eriksson, S. 2010, ApJL 719, L36.}
\ref{Grigis, P., Su, Y., Weber, M., AIA Team
	2012, {\sl AIA PSF Characterization and Image Deconvolution},
	Memo Version 2012-Feb-13.}
\ref{Grinon-Marin, A.B., Socas-Navarro, H., and Centeno, R.
	2017, AA 604, A36.}
\ref{Guerrero, G., Smolarkiewicz, P.K., de Gouveia Dal Pino, E.M.
	et al. 2016, ApJL 828, L3.}
\ref{Higginson, A.K. and Lynch, B.J. 2018, ApJ 859, 6.}
\ref{Howard, R. and LaBonte, B.J. 1980, ApJL 239, L33.}
\ref{Howe,R., Rempel, M., Christensen-Dalsgaard, J., Hill, F.,
	Komm, R., Larsen, R.M., Schou, J., and Thompson, M.J.,
 	2006, ApJ 649, 1155.}
\ref{Howe, R., Christensen-Dalsgaard, J., Hill, F., Komm, R.,
	Schou, J., and Thompson, M.J. 2009, ApJL 701, L87.}
\ref{Howe, R., Christensen-Dalsgaard, J., Hill, F., Komm, R.,
	Larson, T.P., Rempel, M., Schou, J., and Thompson, M.J.
	2013, ApJL 767, L20.}
\ref{Karami, K., and Bahari, K. 2011, Astrophys.Spac.Sci. 333, 463.}
\ref{Kitchatinov, L.L., Pipin, V.V., Makarov, V.I. and Tlatov, A.G.
	1999, SoPh 189, 227.}
\ref{Kosovichev, A.G. and Pipin, V.V. 2019, ApJL 871, L20.}
\ref{Lanza, A.F. 2007, AA 471, 1011.}
\ref{Lemen, J.R., Title, A.M., Akin, D.J., Boerner, P.F., 
	Chou, C., Drake, J.F., Duncan, D.W., Edwards, C.G., et al.
 	2012, SoPh 275, 17.}
\ref{Lekshmi, B., Nandy, D., and Antia, H.M. 2018, ApJL 861, 121L.}
\ref{Link, B. and van Eysden, C.A. 2016, ApJL 823, L1.}
\ref{Luo, Q.Y., Wei, F.S., and Feng, X.S. 2002, SoPh 205, 39.}
\ref{Morton, R.J., Ruderman, M.S. and Erdelyi, R. 2011, AA 534, A27.}
\ref{Mozafari Ghorava A. and Vasheghani-Farahani, S. 2018, ApJ 869, 93.}
\ref{Murawski, K., Solovev, A., Musielak, Z.E., Srivastava, A.K.,
	and Kraskiewicz, J. 2015, AA 577, A126.}
\ref{Nakariakov, V.M., Ofman, L., DeLuca, E., Roberts, B., and
	Davila, J.M. 1999, Science 285, 862.}
\ref{Narain, U., Agarwal, P., Sharma, R.K., Prasad, L. and
	Dwivedi, B.N. 2001, SoPh 199, 307.}
\ref{Nistico, G., Nakariakov, V.M., and Verwichte, E. 2013,
        AA 552, A57.}
\ref{Noble, M.W., Musielak, Z.E., and Ulmschneider, P.
	2003, AA 409, 1085.}
\ref{Pesnell, W.D., Thompson, B.J., and Chamberlin, P.C.
        2011, SoPh 275, 3.}
\ref{Petrovay, K. and Forgacs-Dajika, E. 2002, SoPh 205, 39.}
\ref{Poduval, B., DeForest, C.E., Schmelz, J.T., and Pathak, S.
	2013, ApJ 765:144.}
\ref{Raghav, A.N., Kule, A., Bhaskar, A., Mishra, W., Vichare, G.
	and Surve S. 2018a, ApJ 860, 26.}
\ref{Raghav, A.N. and Kule, A. 2018, MNRAS 476, L6.}
\ref{Rempel, M. 2006, ApJ 647, 662.}
\ref{Rempel, M. 2007, ApJ 655, 651.}
\ref{Rempel, M. 2012, ApJL 750, L8.}
\ref{Roberts, B., Edwin, P.M., and Benz, A.O. 1984, ApJ 279, 857.}
\ref{Roberts, B. 2019, {\sl MHD waves in the solar atmosphere},
	Cambridge University Press.}
\ref{Routh, S., Musielak, Z.E., and Hammer, R. 2007, SoPh 246, 133.}
\ref{Routh, S., Musielak, Z.E., and Hammer, R. 2010, ApJ 709, 1297.}
\ref{Ruderman, M.S., and Roberts, B. 2002, ApJ 577, 475.}
\ref{Sakai, J.I., Minamizuka, R., Kawata, R., and Cranmer, N.F.
	2001, ApJ 550, 1075.}
\ref{Sandman, A., Aschwanden, M.J., DeRosa, M., Wuelser, J.P., Alexander,D.
 	2009, SoPh 259, 1.}
\ref{Sandman, A.W. and Aschwanden, M.J.
 	2011, SoPh 270, 503.}
\ref{Scherrer, P.H., Schou, J., Bush, R.I., Kosovichev, A.G., 
	Bogart, R.S., Hoeksema, J.T., Liu, Y., Duvall, T.L., et al.
 	2012, SoPh 275, 207.}
\ref{Sekse, D.H., Rouppe van der Voort, L., De Pontieu, B.,
	and Scullion, E. 2013, ApJ 769, 44.}
\ref{Shelyag, S. 2015, ApJ 801, 46.}
\ref{Snow, B., Fedun, V., Gent, F.A., Verth, G., and Erdelyi, R.
	2018, ApJ 857, 125.}
\ref{Soler, R., Terradas, J., Oliver, R., and Ballester, J.L.
	2017. ApJ 840, 20.}
\ref{Soler, R., Terradas, J., Oliver, R., and Ballester, J.L.
	2019, ApJ 871, 3.}
\ref{Spruit, H.C. 2003, SoPh 213, 1.}
\ref{Strohmayer, T.E. and Watts, A.L. 2006, ApJ 653, 593.}
\ref{Tapping K.F. 1983, SoPh 87, 177.}
\ref{Vasheghani-Farahani, S., Nakariakov, V.M., and van Doorsselaere, T.
	2010, AA 517, A29.}
\ref{Vasheghani-Farahani, S., Nakariakov, V.M., van Doorsselaere, T.,
	and Verwichte, E. 2011, AA 526, A80.}
\ref{Vasheghani-Farahani, S., Ghanbari, E., Chaffari, G., and Safari, H.
	2017, AA 599, A19.}
\ref{Verth, G., Erdelyi, R., and Goossens, M. 2010, ApJ 714, 1637.}
\ref{Vorontsov, S.V., Christensen-Dalsgaard, J., Schou, J.,
	Strakhov, V.N. and Thompson M.J.
	2002, Science, 296, 101.}
\ref{Wang, T.J., Solanki, S.K., Innes, D.E., et al.
	2003a, AA 402, L17.}
\ref{Wang, T.J., Solanki, S.K., Curdt, W. et al.
	2003b, AA 406, 1105.}
\ref{Warren, H.P., Crump, N.A., Ugarte-Urra,I., Sun,X., Aschwanden,M.J., 
	and Wiegelmann,T. 2018, ApJ 860, 46.}
\ref{Wiegelmann, T., Inhester, B., and Sakurai, T. 2006, 
 	SoPh 233, 215.}
\ref{Wojcik, D., Murawski, K., Musielak, Z.E., Konkol, P., and Mignone, A. 2017, SoPh 292, 31.}
\ref{Zaqarashvili, T.V. 2003, AA 399, L15.}
\ref{Zaqarashvili, T.V., Oliver, R., and Ballester, J.L. 2006, AA 456, L13.}
\ref{Zaqarashvili, T.V. and Murawski, K. 2007, AA 470, 353.}
\ref{Zaqarashvili, T.V. and Belvedere, G. 2007, ApJ 663, 553.}
\ref{Zaqarashvili, T.V., Khodachenko, M.L., and Solar, R.
	2013, AA 549, A113.}
\ref{Zhao, J. and Kosovichev, A.G. 2004, ApJ 603, 776.}

%% REFERENCES

\clearpage

%%%%%%%%%%%%%%% TABLE %%%%%%%%%%%%%%%%%%%%%%%%%%%%%%%%%%%
\begin{table}
\tabletypesize{\normalsize}
\caption{Results of oscillation analysis 
(D=flare duration (min), $n_p$=number of pulses, 
CCC=cross-correlation coefficient, $q_{elim}$=fraction of eliminated loop segments;
P=Oscillation time period in minutes, $A_0$=Initial free energy level 
($10^{30}$ erg), $A_1$=linear drift of free energy level ($10^{30}$ erg hour$^{-1}$), 
$A_2$=Amplitude of oscillation level ($10^{30}$ erg), M=modulation depth, median 
misalignment angle $\mu_2$ of VCA4-NLFFF magnetic field model with respect
to observed loop directions), $L$=length of mean field line of dominant dipole,
and $v_{phase}$=phase speed along mean field line.}
\medskip
\begin{tabular}{rrrrrrrrrrrrrrr}
\hline
\#  & Date     & Time   & D    &$n_p$& CCC &$q_{elim}$ &P &$A_0$  &  $A_1$ &  $A_2$ &    M & $\mu_2$ & $L$ & $v_{phase}$ \\
    &          & [UT]   & (min)&    &      &      & (min) &       &        &        &      & (deg)   & (Mm) & (km/s)     \\
\hline
\hline
 12 & 20110215 & 023800 &   24 &  2 & 0.82 & 0.70 &  11.3 &  0.51 &   0.16 &  -0.09 &   0.17 &  9.4  & 100 & 300 \\
 37 & 20110310 & 000100 &   17 &  1 & 0.80 & 0.70 &  15.2 &  0.34 &   0.08 &   0.04 &   0.12 &  5.3  & 110 & 250 \\
 66 & 20110906 & 225700 &   15 &  1 & 0.83 & 0.80 &  15.4 &  0.86 &  -0.49 &  -0.12 &   0.14 &  7.2  & 160 & 350 \\
 67 & 20110907 & 231700 &   15 &  1 & 0.90 & 0.70 &  14.2 &  0.82 &  -0.50 &   0.28 &   0.35 &  7.1  & 170 & 400 \\
147 & 20120307 & 010800 &   36 &  2 & 0.77 & 0.70 &  19.8 &  1.00 &  -0.29 &   0.13 &   0.13 & 14.7  & 130 & 220 \\
148 & 20120307 & 015600 &   20 &  2 & 0.88 & 0.90 &  13.4 &  1.37 &  -0.47 &  -0.12 &   0.09 &  5.7  &  70 & 180 \\
220 & 20120712 & 173600 &   21 &  1 & 0.89 & 0.70 &  22.5 &  2.29 &   2.78 &   0.73 &   0.32 & 14.5  & 100 & 150 \\
344 & 20131105 & 224600 &    8 &  1 & 0.76 & 0.90 &  12.3 &  1.25 &   0.42 &  -0.53 &   0.42 &  8.8  & 130 & 350 \\
349 & 20131108 & 050200 &   12 &  1 & 0.67 & 0.90 &  10.0 &  1.43 &   0.09 &   0.12 &   0.08 & 13.5  & 170 & 570\\
351 & 20131110 & 055000 &   12 &  1 & 0.89 & 0.70 &  19.2 &  0.57 &   0.54 &   0.16 &   0.29 &  8.0  & 200 & 340 \\
384 & 20140107 & 193100 &   56 &  4 & 0.83 & 0.60 &  12.3 &  1.43 &   0.76 &   0.32 &   0.22 & 11.9  & 130 & 350 \\
\hline
mean&          &        &   21 &    & 0.83 &      & 15.1 &       &        &        &        &       & 135    & 315    \\
    &          &        &$\pm14$&   &$\pm0.07$&   &$\pm3.9$&     &        &        &        &       &$\pm35$ &$\pm120$\\
\hline
\end{tabular}
\end{table}

\clearpage

%%%%%%%%%%%%%%% FIGURES %%%%%%%%%%%%%%%%%%%%%%%%%%%%%%%%%%%

\begin{figure}
\centerline{\includegraphics[width=1.0\textwidth]{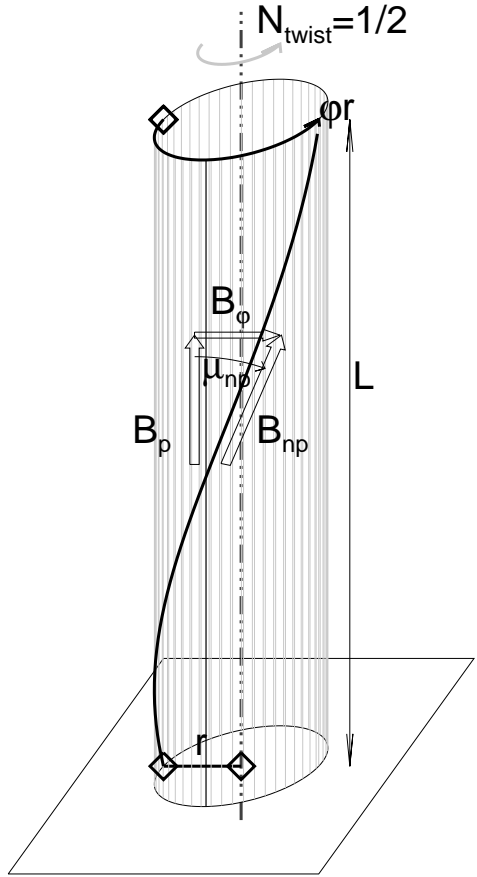}}
\caption{Geometric concept of the vertical-current approximation model
of a torsionally twisted flux tube:
The potential field component $B_r=B_p$ is aligned with the
flux tube symmetry axis, the nonpotential field component 
$B_{np}$ follows the torsionally twisted flux tube with a
constant misalignment angle $\mu_{np}$, and the azimuthal 
field component $B_{\varphi}$ is orthogonal to the potential
field component $B_p$.  The twist corresponds to a half turn 
($n_{twist}=0.5$) over a flux tube length $L$ at a radius $r$.}
\end{figure}

\begin{figure}
\centerline{\includegraphics[width=1.0\textwidth]{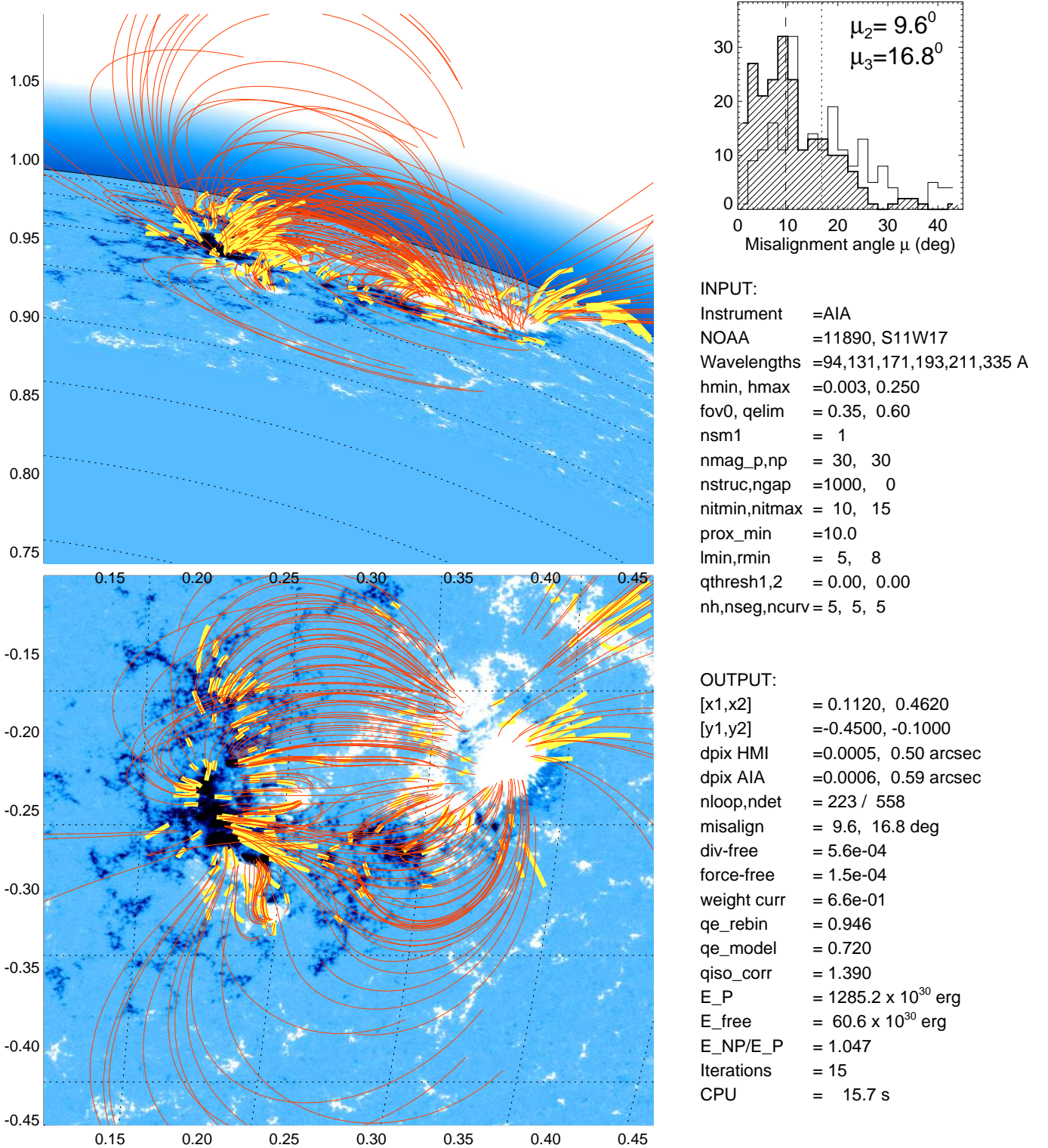}}
\caption{HMI magnetogram (blue), automatically traced coronal
loop segments in AIA (yellow), and magnetic field lines computed with the
VCA4-NLFFF code (red curves) for flare event \# 351, observed
on 2013-Nov-10 04:38:00 UT. The line-of-sight magnetogram
$B_z(x,y)$ is shown (bottom panel), as well as rotated to the
north by $90^\circ$ (top panel). Histograms of misalignment
angles are shown for the 2-D projected ($\mu_2=9.6^\circ$) and the
3-D reconstructed ($\mu_3=16.8^\circ$) loop directions with respect
to the VCA4-NLFFF magnetic field model.}
\end{figure}

\begin{figure}
\centerline{\includegraphics[width=1.0\textwidth]{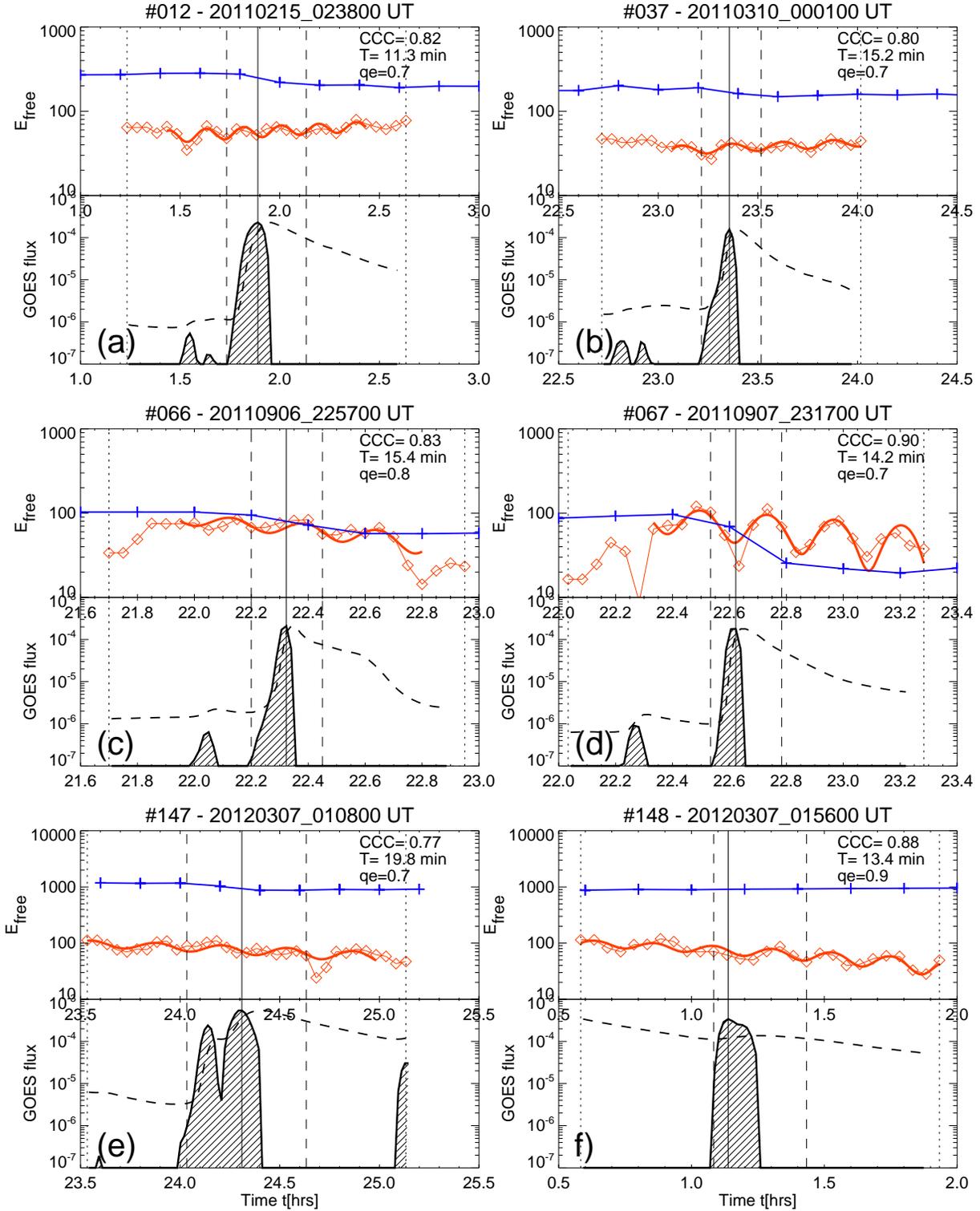}}
\caption{Time evolution of the free energy $E_f(t)$ (red diamonds),
fit of oscillatory function (red thick curve), free energy 
evolution from Wiegelmann NLFFF code (blue curves),
GOES light curves at 1-8 \ang\ (dashed black curves),
and time derivative of GOES light curve (hashed areas).
The flare start and end times (dashed vertical lines) 
according to NOAA.}
\end{figure}

\begin{figure}
\centerline{\includegraphics[width=1.0\textwidth]{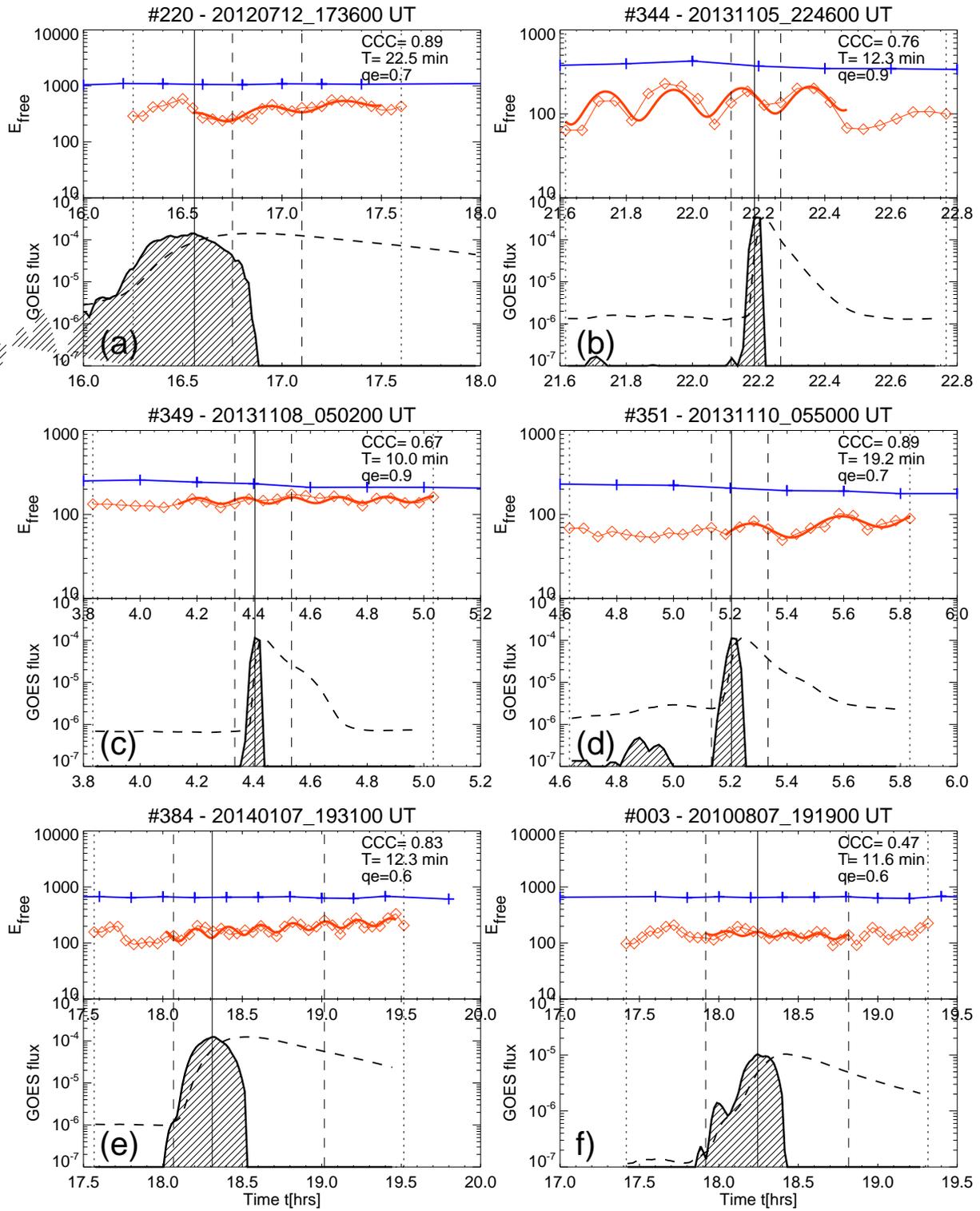}}
\caption{Similar representation as in Fig.~4 for 5 more 
X-ray flares.}
\end{figure}

\begin{figure}
\centerline{\includegraphics[width=1.0\textwidth]{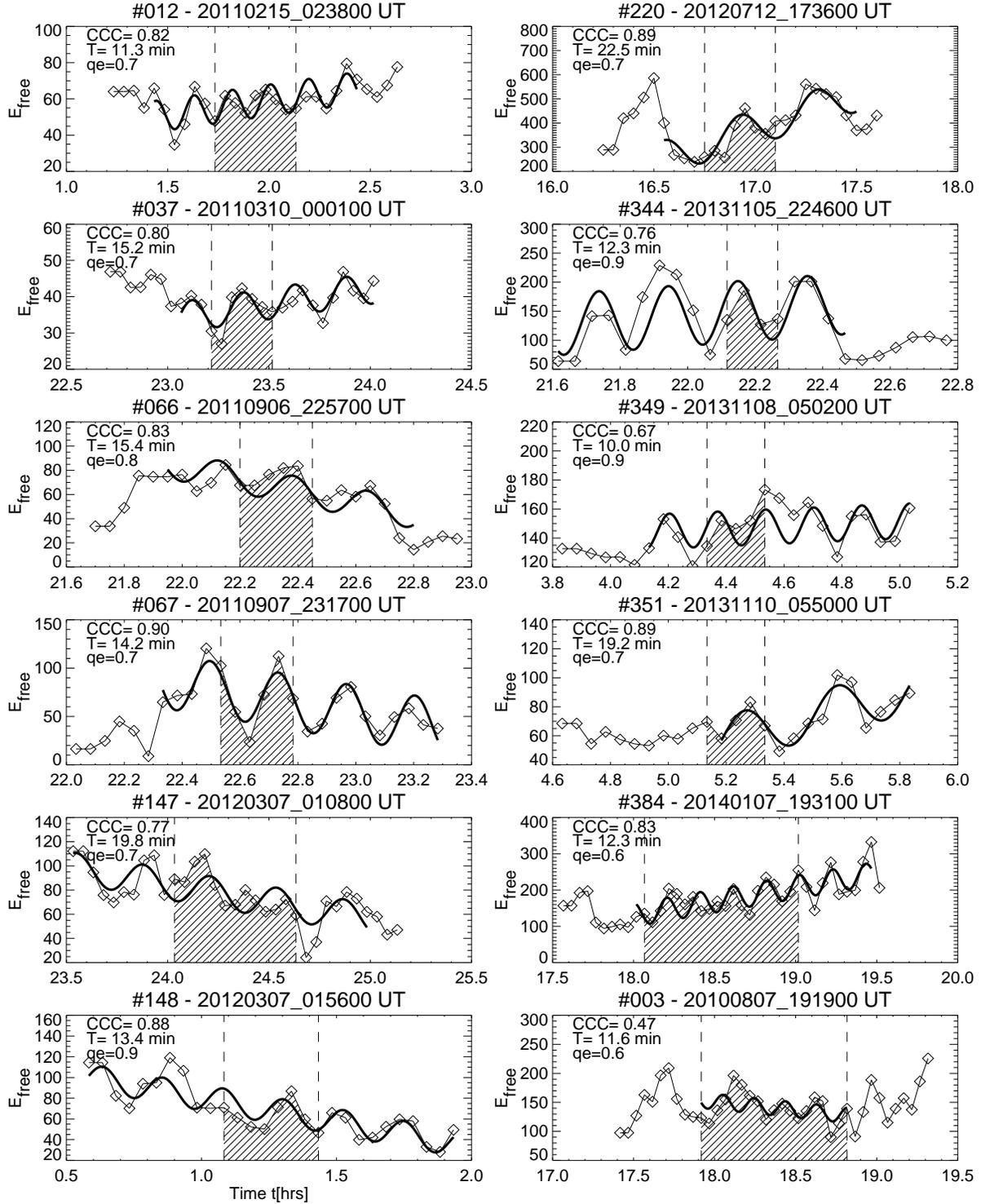}}
\caption{The free energy $E_f(t)$ as a function of the time
is shown for 11 X-class flares (diamonds), together with a
fit of a sinusoidal modulation function (thick black curves).
The cadence is $\Delta t=3$ minutes, bracketed by the flare
start and end times (vertical dashed lines) as defined in 
GOES data by NOAA.}
\end{figure}

\begin{figure}
\centerline{\includegraphics[width=1.0\textwidth]{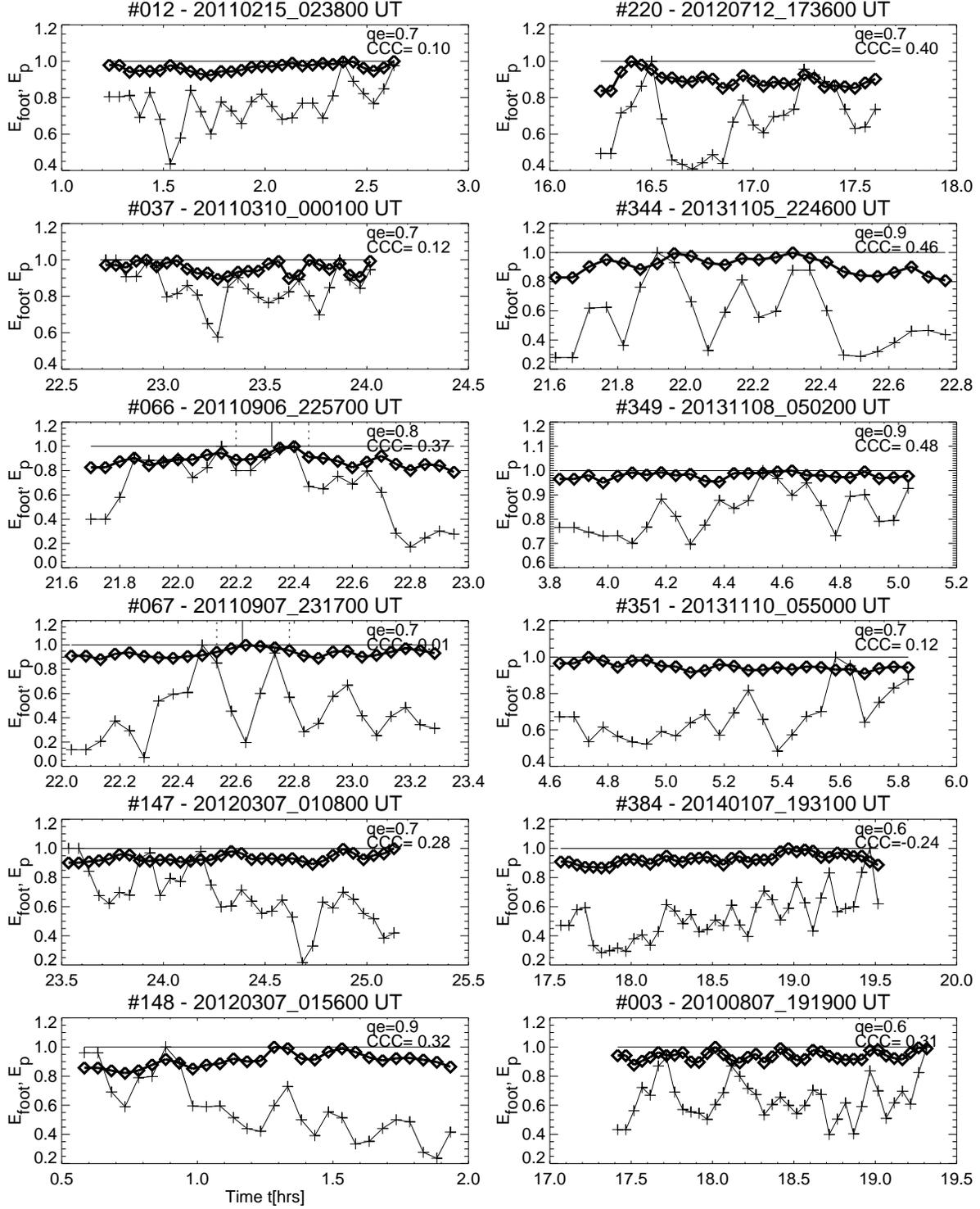}}
\caption{Time evolution of the potential field energy $E_p(t)$
(thick curve with diamonds) and the free energy
$E_{f}(t)$ (thin curve with crosses), both normalized to
their maximum, the loop elimination ratio $q_e$, and 
cross-correlation coefficient $CCC$. Note that the potential 
field energy and the free energy are uncorrelated.}
\end{figure}

\begin{figure}
\centerline{\includegraphics[width=0.7\textwidth]{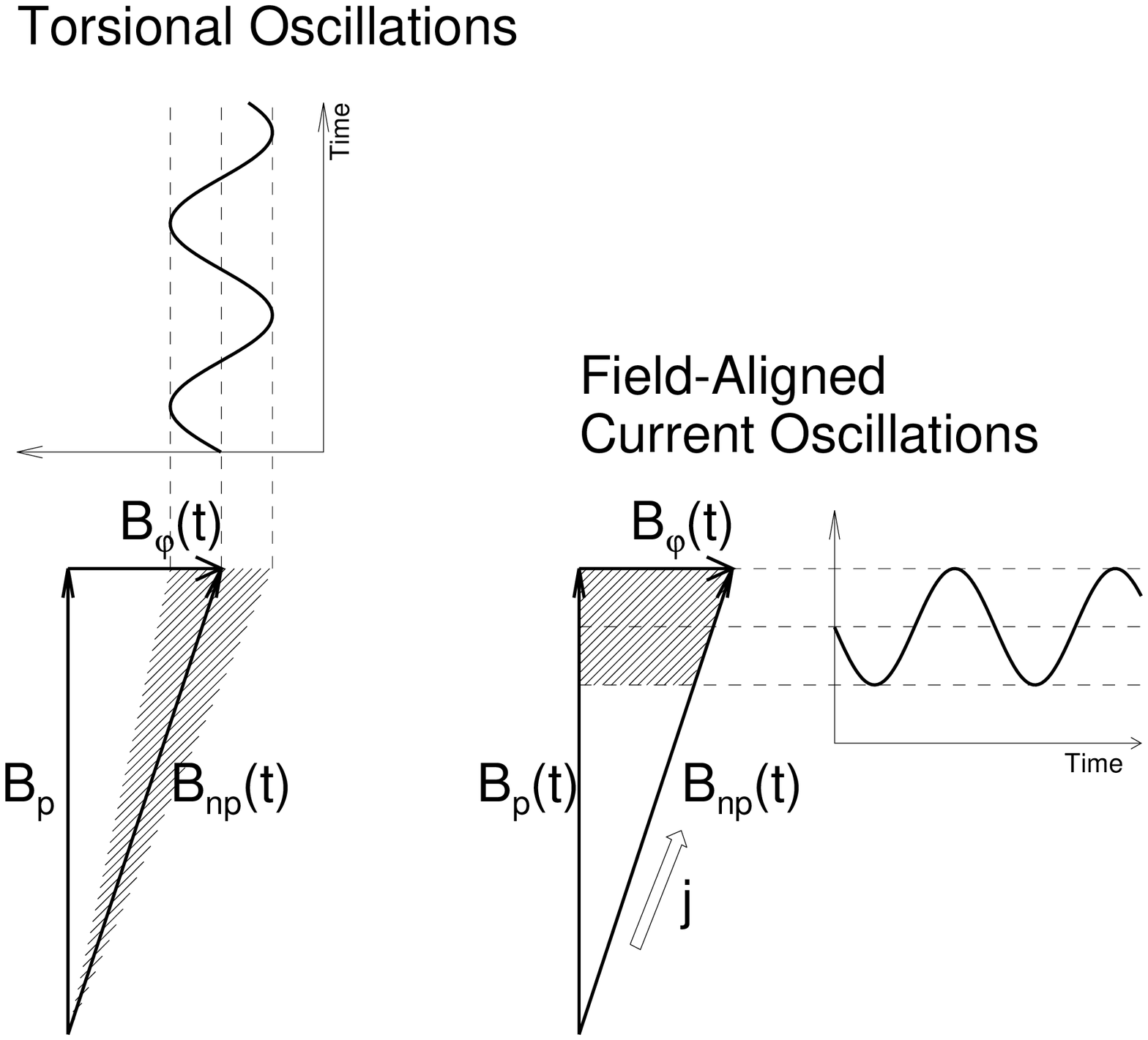}}
\caption{Two scenarios of oscillating magnetic field components:
torsional oscillations with $B_p=const$ and $B_{\varphi}$
oscillating (left side); and field-aligned current oscillations,
where the current ${\bf j} \propto (\nabla \times {\bf B})$
oscillates, modulating all components $B_p$, $B_{np}$,
and $B_{\varphi}$ proportionally (right side).}
\end{figure}

\begin{figure}
\centerline{\includegraphics[width=1.0\textwidth]{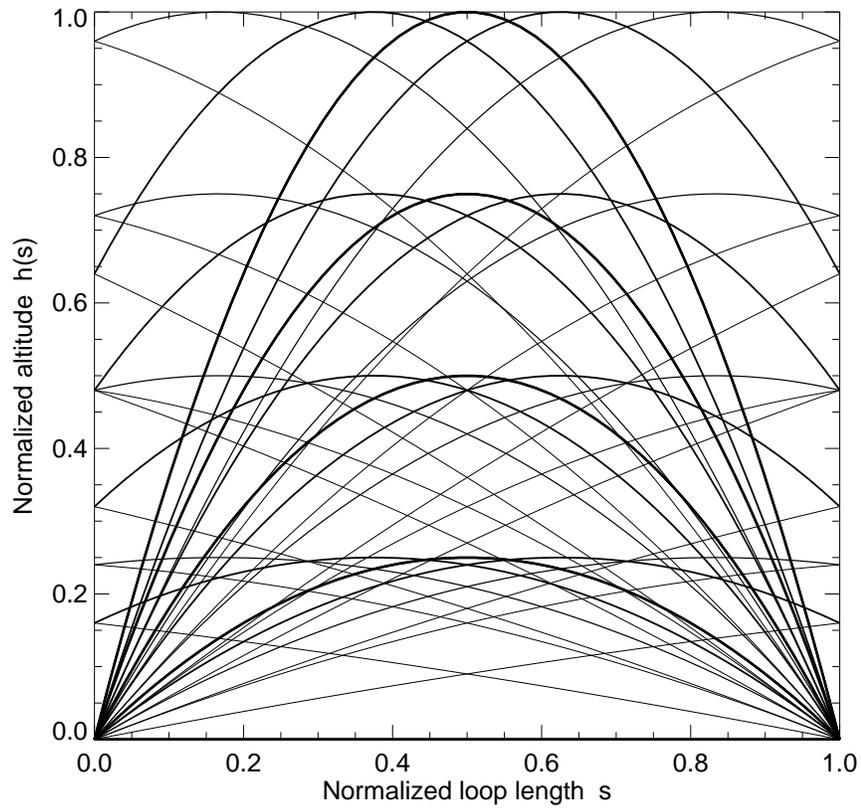}}
\caption{Trial geometries for segments of loop altitudes $h(s)$.}
\end{figure}

\end{document}